%% file: main_tr.tex
\documentclass[preprint,10pt,5p]{elsarticle}
\input{preamble}

\begin{document}

\begin{frontmatter}
\title{Computational Nuclear Quantum Many-Body Problem: The UNEDF Project}

\author[msu]{S. Bogner}
\author[uw]{A. Bulgac}
\author[lanl]{J. Carlson}
\author[unc]{J. Engel}
\author[ornlc]{G. Fann}
\author[osu]{R.J.~Furnstahl}
\author[lanl]{S. Gandolfi}
\author[ornlp]{G. Hagen}
\author[cmu]{M.~Horoi}
\author[sdsu]{C.~Johnson}
\author[ut,ornlp,uj]{M. Kortelainen}
\author[anl]{E. Lusk}
\author[isu]{P. Maris}
\author[ornln]{H. Nam}
\author[llnl,triumf]{P. Navratil}  
\author[ut,ornlp,uwpl]{W. Nazarewicz}
\author[lbnl]{E. Ng}
\author[bnl]{G.P.A.~Nobre}
\author[llnl]{E. Ormand}
\author[ornlp,ut]{T. Papenbrock}
\author[ut,ornlp,pku]{J. Pei} 
\author[anlp]{S. C. Pieper}
\author[llnl]{S. Quaglioni}
\author[pnnl,uw]{K.J. Roche}
\author[anl]{J. Sarich}
\author[llnl]{N.~Schunck}
\author[odu]{M. Sosonkina} 
\author[tsukuba,unc]{J.~Terasaki}
\author[llnl]{I. Thompson}
\author[isu]{J.P.~Vary}
\author[anl]{S.M.~Wild}

\address[anl]{Mathematics and Computer Science Division, Argonne National
Laboratory, Argonne, IL 60439, USA}
\address[anlp]{Physics Division, Argonne National Laboratory, Argonne, IL 60439,
USA}
\address[bnl]{National Nuclear Data Center, Brookhaven National Laboratory,
Upton, NY 11973, USA}
\address[cmu]{Central Michigan University, Mount Pleasant, MI 48859, USA}
\address[isu]{Department of Physics and Astronomy, Iowa State University, Ames,
IA 50011, USA}
\address[lanl]{Theoretical Division, Los Alamos National Laboratory, Los Alamos,
NM 87545, USA}
\address[lbnl]{Computational Research Division, Lawrence Berkeley National
Laboratory, Berkeley, CA 94720, USA}
\address[llnl]{Physics Division, Lawrence Livermore National
Laboratory, Livermore, CA 94551, USA}
\address[msu]{National Superconducting Cyclotron Lab, Michigan State
University, East Lansing, MI, 48824, USA}
\address[ornlc]{Computer Science and Mathematics Division, Oak Ridge National
Laboratory, Oak Ridge, TN 37831, USA}
\address[ornln]{National Center for Computational Sciences Division, Oak
Ridge National Laboratory, Oak Ridge, TN 37831, USA}
\address[ornlp]{Physics Division, Oak Ridge National Laboratory, Oak Ridge, TN
37831, USA}
\address[osu]{Department of Physics, Ohio State University, Columbus, OH 43210,
USA}
\address[odu]{Department of Modeling, Simulation and Visualization Engineering,
Old Dominion University, Norfolk, VA 23529, USA}
\address[pnnl]{Computational Sciences and Mathematics Division, Pacific
Northwest National Laboratory, Richland, WA 99352, USA}
\address[pku] {State Key Laboratory of Nuclear Physics and Technology, School of
Physics, Peking University, Beijing 100871, China}
\address[sdsu]{Department of Physics, San Diego State University, San Diego, CA
92182, USA}
\address[triumf]{TRIUMF, 4004 Westbrook Mall, Vancouver, BC, V6T 2A3, Canada} 
\address[uj]{Department of Physics, P.O. Box 35 (YFL), FI-40014, University of
Jyv\"askyl\"a, Finland}
\address[unc]{Department of Physics and Astronomy, University of North Carolina,
Chapel Hill, NC 27599, USA}
\address[ut]{Department of Physics and Astronomy, University of
Tennessee, Knoxville, TN 37996, USA}
\address[tsukuba]{Division of Physics and Center for Computational Sciences,
University of Tsukuba, Tsukuba, 305-8577, Japan} 
\address[uwpl]{Faculty of Physics, University of Warsaw, 00-681 Warsaw, Poland}
\address[uw]{Department of Physics, University of Washington, Seattle, WA 98195,
USA}


\begin{abstract}

The UNEDF project was a large-scale collaborative effort that applied
high-performance computing to the nuclear quantum many-body problem.  UNEDF
demonstrated that close associations among nuclear physicists, mathematicians,
and computer scientists can
lead to novel physics outcomes built on algorithmic innovations and
computational
developments.  This review showcases a wide range of UNEDF science results to
illustrate this interplay.
 
\end{abstract}


\end{frontmatter}

\section{Introduction to UNEDF}
\label{sec:intro}

Understanding the properties of atomic nuclei is crucial for a complete nuclear
theory, for element formation, for properties of stars, and for present and
future energy and defense applications. From 2006 to 2012, the UNEDF
(Universal Nuclear Energy Density Functional) collaboration carried out a
comprehensive study of the nuclear many-body problem using advanced numerical
algorithms and extensive computational resources, with a view toward scaling
to petaflop supercomputing platforms and beyond. 

The UNEDF project was carried out as part of the SciDAC (Scientific
Discovery through Advanced Computing) program led by
Advanced Scientific Computing Research (ASCR), part of the Office of Science
in the U.S. Department of Energy (DOE). 
The SciDAC program was
started in 2001 as a way to couple the applied mathematics and computer
science research sponsored by ASCR to applied computational science
application projects traditionally supported by other offices in DOE.
UNEDF was funded jointly by ASCR, the Nuclear
Physics program of the Office of Science, and the National Nuclear
Security Administration.
Over 50 physicists, applied mathematicians, and computer scientists from 9
universities and 7 national laboratories in the United States, as well as  many
international collaborators, participated in UNEDF. 

This review describes science outcomes in nuclear many-body physics, with an
emphasis on
computational and algorithmic developments,  that have
resulted from the successful collaborations within UNEDF among mathematicians
and
computer scientists on one side and nuclear physicists on the other.
Such 
collaborations ``across the divide'' were newly formed at the early stage of the
project and became its unique feature, with
high-performance computing serving as a catalyst for new interactions.
The results described in this paper could not have
been achieved without such couplings.

\subsection{UNEDF science}

The long-term vision initiated with UNEDF is to arrive at a
comprehensive, quantitative, and unified description of nuclei and their
reactions that is grounded in the fundamental interactions between the
constituent nucleons \cite{Bertsch:2007,Furn:2011}. The goal is to replace 
phenomenological models of nuclear structure and reactions with a well-founded
microscopic theory that delivers maximum predictive power with well-quantified
uncertainties. Specifically, the mission of UNEDF was threefold:
\begin{enumerate}
 \item Find an optimized energy density functional (EDF) using all our
knowledge of the nucleonic Hamiltonian and basic nuclear properties. 
 \item Validate the functional using  the  relevant nuclear  data. 
 \item Apply the validated theory to properties of interest that 
cannot be measured.
\end{enumerate}

\begin{figure}[tb!]
\includegraphics[width=\linewidth]{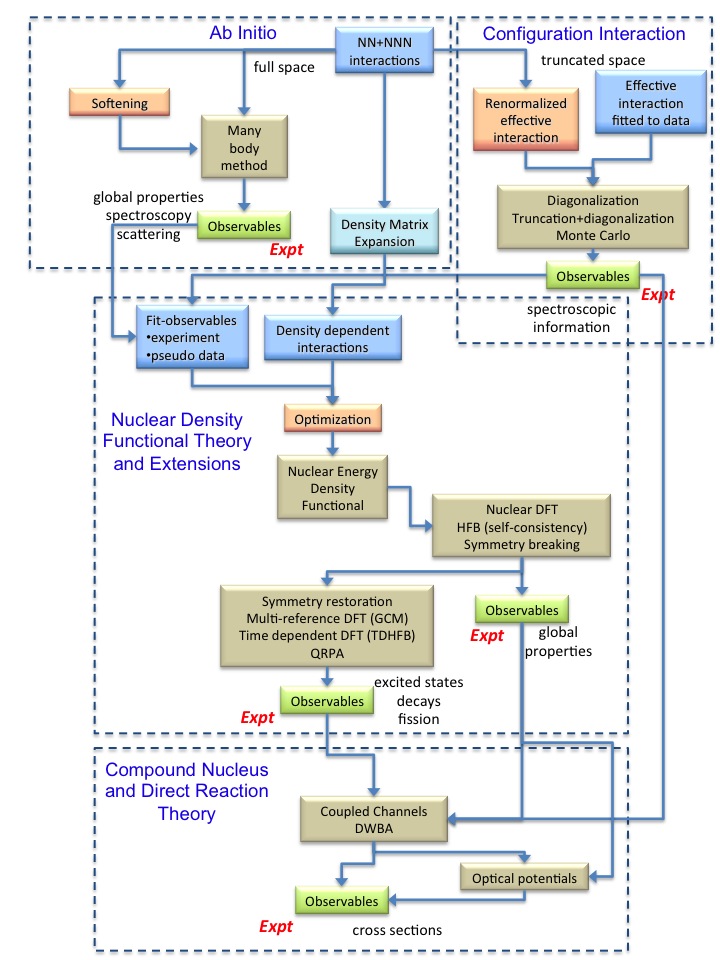}
\caption{UNEDF project scope.  Major science  areas are indicated by boxes; 
interconnections between areas are marked by arrows.  The green boxes
indicate connections to experimental observations.} 
\label{fig:unedf}
\end{figure}

The main physics areas of UNEDF, defined at the beginning of the project
\cite{Bertsch:2007}, were
ab initio structure, ab initio functionals, density functional theory
(DFT) applications, DFT extensions, and reactions. Few connections between
these areas existed at that time. As UNEDF matured, however, coherence
grew within the effort. Indeed, the project created and facilitated an
increasing interplay among the major areas where none had existed previously.
Each of the main physics areas now includes ongoing collaborations that cross
over into other areas. These interconnections are highlighted in the summary
diagram of the UNEDF strategy shown in \Fig{fig:unedf}. In addition to
physics links, numerous computer science/applied mathematics (CS/AM)
interconnections were established within UNEDF as computational and
mathematical tools developed in one area of
UNEDF were used in other parts of the project. These tools, motivated by
nuclear needs, are now available for other areas of science.
Access to leadership-class computing resources and large-scale compute time
allocations were critical for the scientific investigations.  

At the intersection of the ab initio techniques and DFT techniques are
comparisons of observables among the various approaches, particularly through
constraints on density. Such calculations have not been performed before and
require significant computational capability and an increasing sophistication of
data manipulation.  Research on the nuclear problem would be incomplete without
a serious effort to understand the nuclear interactions involved and their
connection to DFT. Therefore, the UNEDF project also included elements that
required less computational capability but are integral to the project,
such as the development of nuclear forces using renormalization group
approaches. Another example is research on nuclear reaction properties that
requires both the use and development of algorithms for the largest computers
and more conventional computing needed for  algorithmic breakthroughs. 

Another new aspect of the nuclear theory effort driven by this project is a
greatly 
enhanced degree of quality control.  Integral to UNEDF was the
verification of methods and codes, the estimation of uncertainties, and
other output assessments. Methods used for verification and validation included
the crosschecking of 
different theoretical methods and codes, the use of multiple DFT solvers with
benchmarking, and benchmarking of different ab initio methods using the same
Hamiltonian.  A new way to estimate theory error bars
was to use multiple Hamiltonians with different energy/momentum cutoffs and then
analyze the
cutoff dependence of calculated observables. The UNEDF assessment component
necessitated the development and application of statistical tools to deliver
uncertainty quantification and error analysis for theoretical studies as well as
to assess the significance of new experimental data.  
Such technologies are essential as new theories and computational tools are
applied to entirely new nuclear systems and to conditions that are not
accessible to experiment.

\subsection{Collaborative effort}

The successes of the UNEDF project were built upon certain best
practices, some implemented originally and some learned by experience, in
organizing and implementing the scientific effort. 
In order to foster the close alignment of the necessary applied mathematics and
computer
science research with the necessary physics research, multiple direct
partnerships were formed consisting of computer scientists and applied
mathematicians linked with specific physicists  to remove algorithmic and/or
computational barriers to progress. The five-year lifetime of the project
provided time for these collaborations to become deep, and they have
continued into follow-up projects.

All these partnerships have success stories to tell, from greatly improved
load balancing on leadership-class machines, to new DFT solver technologies, to
dramatically improved algorithms 
for optimization of functionals, to eigenvalues and eigenfunctions of
extremely large matrices, and more. 

The SciDAC program aims at transformative science, and this goal has been
fulfilled by the new capabilities stemming from UNEDF.  
But the outcomes reach
beyond the many compelling nuclear physics calculations. UNEDF has
changed for the better the way that low-energy nuclear theory is carried out,
analogous to the shift in experimental programs, moving from many small groups
working independently to large-scale collaborative  efforts.

\section{Science \pagebudget{}}
\label{sec:science}

\begin{figure}[tb!]
\includegraphics[width=0.99\linewidth]{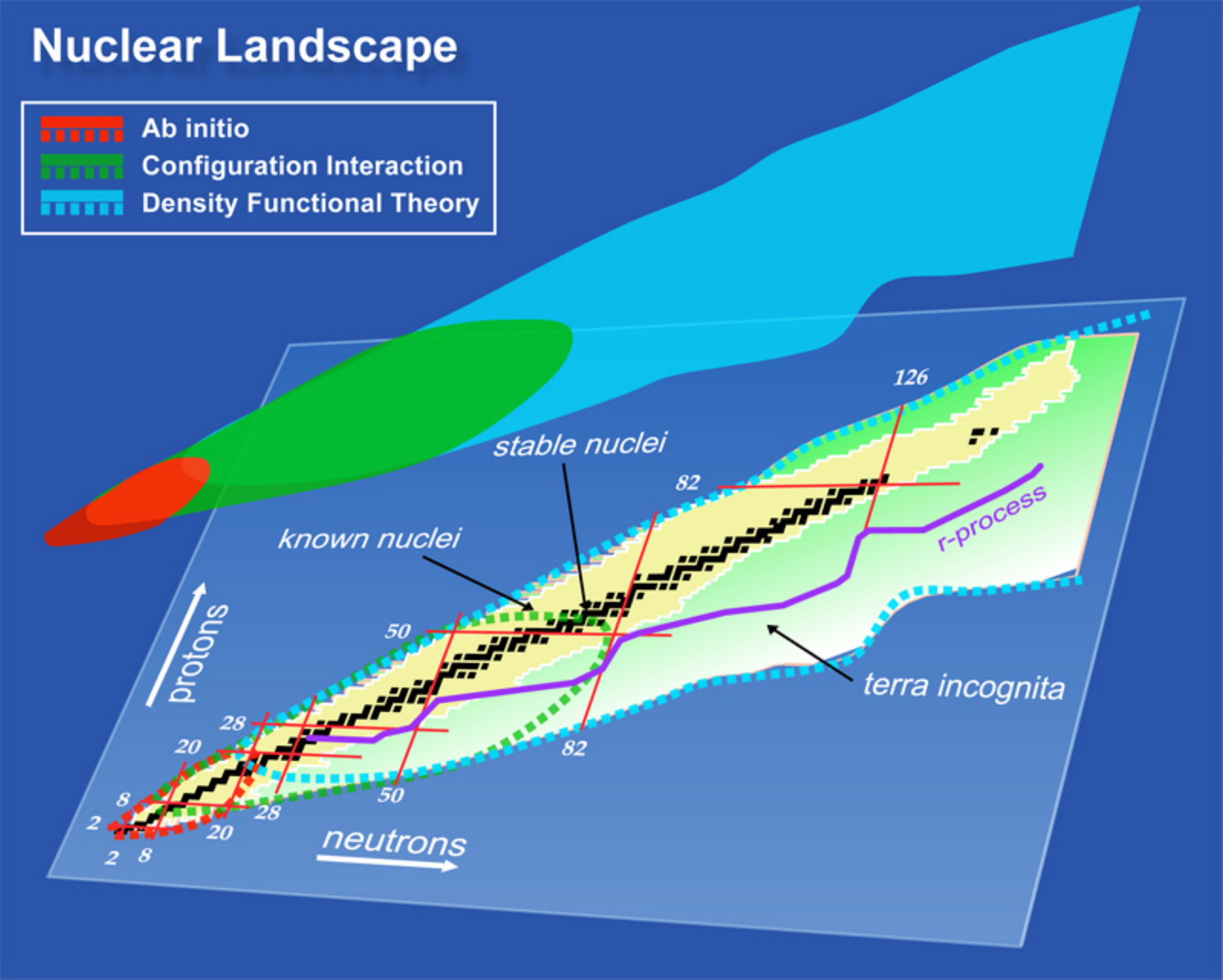}
\caption{Theoretical approaches for solving the nuclear quantum many-body
problem used by UNEDF.  The lightest nuclei can be computed by using ab initio
methods based on the bare internucleon interactions (red). Medium-mass nuclei
can be treated by configuration interaction techniques  (green). For heavy
nuclei, the density functional theory based on the optimized energy density
functional is the tool of choice. (From \cite{Bertsch:2007}.)} 
\label{fig:landscape}
\end{figure}

The territory of UNEDF science  is the chart of the nuclides in the
$(N,Z)$-plane shown in
Fig.~\ref{fig:landscape}. On this chart, stable nuclei are represented  by black
squares, while the yellow squares indicate unstable nuclei that have been seen
in the laboratory. The sizable green area marked ``terra incognita'' is
populated by unstable isotopes  yet to be explored.
Above the table of nuclides are shown three broad classes of theoretical
methods, which are also used in other fields dealing with strongly interacting
many-body systems, such as quantum chemistry  and condensed matter physics.
Light nuclei and their reactions can be computed by using ab initio techniques
(quantum Monte Carlo, no-core shell model) described in
Sec.~\ref{sec:abinitio}.
Medium-mass nuclei can be treated by configuration interaction (CI) techniques
(Sec.~\ref{sec:ci}). The bulk of the nuclides are covered by the nuclear DFT
described in Sec.~\ref{sec:dft}, which provides
the theoretical underpinning and computational framework for building a nuclear
EDF. Time-dependent phenomena involving complex nuclei, including nuclear
reactions, can be described by means of approaches going beyond static DFT
(\Sec{sec:beyonddft}). By enhancing and exploiting the overlaps with ab initio
and
CI approaches, the goal is to
construct and validate a nuclear EDF informed by microscopic interactions as
well as experimental data. 

\subsection{Ab initio methods and benchmarking \pagebudget{(Joe) (1/2
page)}}
\label{sec:abinitio}

 Ab initio methods solve few- and many-body problems by 
using realistic two- and three-nucleon interactions and obtain
the structure and dynamic properties of nuclei. The nuclear interaction
depends on the spatial, spin, and isospin coordinates of the
nucleons. Consequently, calculations are much more computationally
demanding than typical quantum problems.  Items of interest include
nuclear spectra, charge and magnetic ground-state and transition densities,
electron and neutrino scattering, and
low-energy reactions. The main goals are to reproduce known
nuclear properties and predict properties that are difficult or
impossible to measure.

  Several ab initio methods have been developed for studying light nuclei;
all have analogues in the study of  condensed matter and
electronic systems.
Quantum Monte Carlo (QMC) methods, including Green's function Monte Carlo
(GFMC), 
use Monte Carlo evaluations of path integrals, explicitly summing over the
spin states and isospin states of the system. The most recent GFMC calculations
have concentrated on the $^{12}$C nucleus, a fascinating system with
a low-lying excited $0^+$ state, the Hoyle state, very near the threshold
of three-alpha particles. QMC methods have also been used to calculate
the properties of neutron matter and neutrons in inhomogeneous potentials.

  No-core shell model (NCSM) methods, including the large-scale
many-fermion dynamics nuclear (\MFDn) code, expand the interacting states
in products of single-particle states and project the low-lying
states through large-scale matrix operations. \MFDn\ calculations have
been used, for example, to explain the long lifetime of the $^{14}$C
nucleus used in carbon dating.  A combination  of no-core shell model
techniques with the resonating group method is currently used to calculate
important low-energy nuclear reactions.

The coupled-cluster method is an ideal microscopic approach to
describe nuclei with closed (sub)shells and their neighbors. It
exhibits a low computational cost (scales polynomially with system size)
while capturing the dominant parts of correlations in the wave
function. This method has been employed to describe and predict the
structure and reactions of neutron-rich oxygen and calcium isotopes. 

\subsubsection{GFMC \pagebudget{(Rusty, Steve, Joe) (1 page)}}
\label{sec:gfmc}

Green's function Monte Carlo calculations start with an
initial trial state $\Psi_T$ and obtain expectation values 
in the exact eigenfunction $\Psi_0$ of the Hamiltonian.  These calculations 
are done by evolution in imaginary time $\tau$:
$\Psi_0 = \exp [ - H \tau ] \Psi_T$ for sufficiently large $\tau$.  The
evolution is done in
many small steps of $\tau$, each step being a nested $3A$-dimensional 
integral.
GFMC was introduced in light nuclei 
\cite{GFMC1987,Pieper:2001mp} to include the strong correlations
induced by the nuclear interaction.
This method has been used to calculate the spectra of light nuclei
up to $^{12}$C \cite{Pieper:2001mp,GFMCscidac:2010aa}, as well as form factors,
electron scattering, and low-energy
reactions~\cite{Nol:2007}.

Calculations of $^{12}$C require the largest-scale computers available,
using a combination of efficient load-balancing for the Monte Carlo and
large-scale linear algebra for the spin-isospin degrees of freedom.
The calculations of $^{12}$C required the development of 
the Asynchronous Dynamic Load Balancing (\ADLB) library to
efficiently perform the load balancing on more than 100,000 cores
\cite{GFMCscidac:2010aa}.

A program, \AGFMC, has been developed over the past 15 years to
carry out these calculations~\cite{Pud:1997,Wir:2000,Pie:2004}.  
It is a large (80,000 lines) Fortran
code that originally used MPI to manage parallelism.  At the beginning of this
project,
the \AGFMC\ code was scaling well up to around 2,000 processes and
performing satisfactorily on IBM's Blue Gene/L computer.  At that time
it was becoming apparent that if the code were to be able to take advantage of
new, petascale machines expected to come on line during the five-year
project to investigate larger nuclei, a significant increase in the
degree of parallelism would need to be incorporated into its main
algorithms.  The greater degree of parallelism (from thousands to tens
of thousands of processes) would give rise to load-balancing problems
that would strain the then-used approach.

One of the goals of UNEDF was
to construct a software library, intrinsically general-purpose but
with features driven by the requirements of \AGFMC, to attack the
load-balancing problem.  The purposes of the library were to supply a
programming interface that would enable relatively straightforward
migration of the existing \AGFMC\ code to the new load-balancing library
and to scale the entire system to much larger degrees of parallelism.

The result is the \ADLB\ library \cite{GFMCscidac:2010aa}.  \ADLB\ 
generalizes the classical 
manager-worker parallel programming model by allowing application
processes (workers) to \emph{put} arbitrary independent work units into
a shared pool and \emph{get} them out to complete them, notifying other
processes when they have done so.  Work units are assigned types and
priorities by the workers and retrieved according to these properties,
allowing complex algorithms to be implemented, despite the simple nature
of the parallel programming model.  Scalability is achieved by
dedicating a small percentage (but still potentially a large number)
of the job's processes to maintaining this work pool and responding to
\emph{put} and \emph{get} requests.  These ``server'' processes execute
independently from the application processes, thus allowing asynchronous load
balancing of process load, memory consumption for the work pool, and
message traffic.

\begin{figure}[tb]
\includegraphics[width=0.95\linewidth]{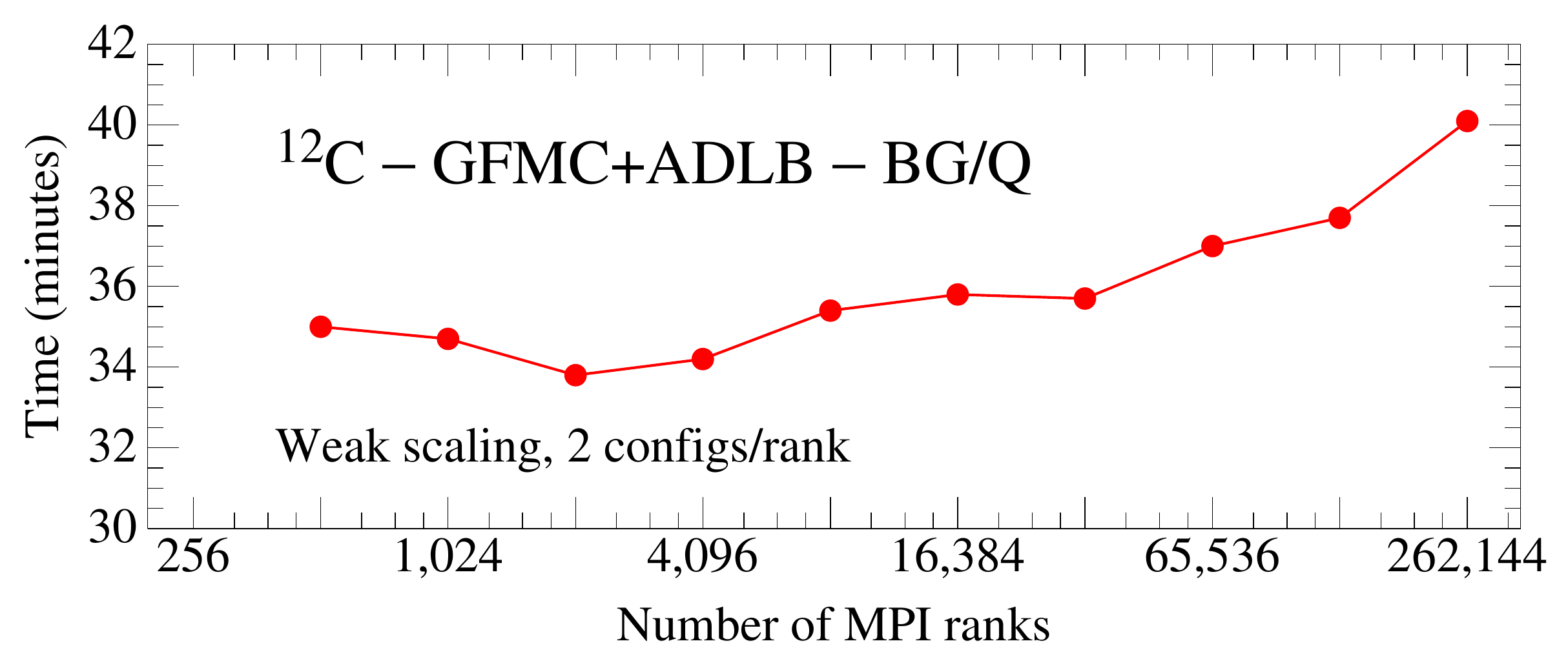}
\caption{Weak scaling of \AGFMC\ with \ADLB\ 
in terms of MPI ranks. There are 8 ranks per BG/Q node; each rank is
using 6 OpenMP threads. Note the compressed vertical scale.}
\label{fig:adlb}
\end{figure}

This scheme has worked well.  Most of the MPI programming in the
original \AGFMC\ code has been absorbed into the \ADLB\ library, yet the
overall code structure has been maintained.  Scalability has been extended
to more than 32,000 processes on BG/P and more than 260,000
processes on BG/Q (see \Fig{fig:adlb}), enabling
scientific results unattainable before this project was undertaken. 

The $^{12}$C nucleus is particularly intriguing because it has a low-lying
$0^+$ excited state (the ``Hoyle'' state) very near the energy of the
breakup into  three alpha particles.  This state is essential for the
nucleosynthesis of carbon in stars through the triple-alpha process.
For $^{12}$C the $\Psi_T$ are linear combinations of shell-model
and alpha-cluster states.
\begin{figure}[htb]
\includegraphics[width=0.95\linewidth]{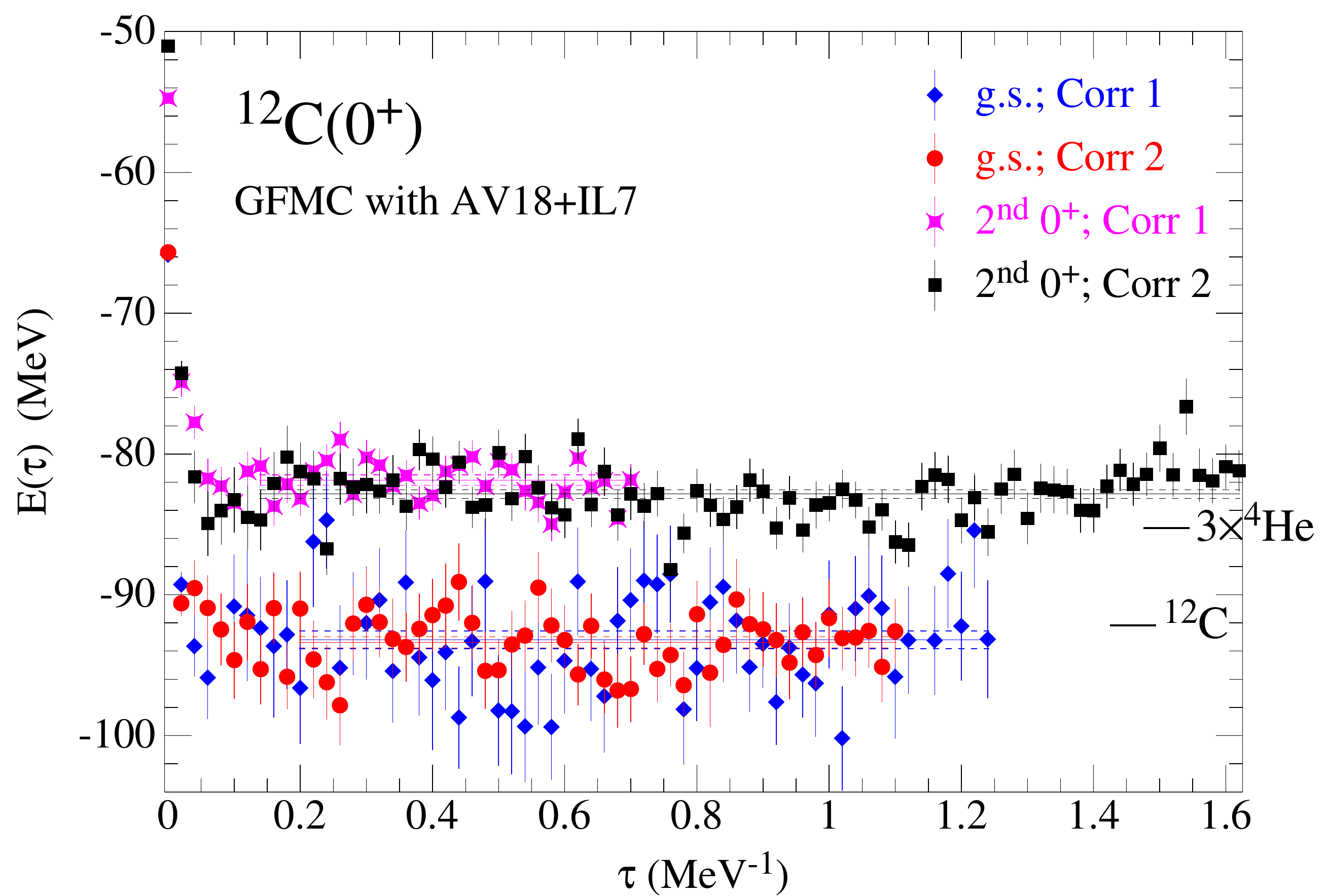}
\caption{Convergence of the ground state (lower curves) and Hoyle state
(upper curves) for different initial states as a function of imaginary time.}
\label{fig:gfmc}
\end{figure}
Figure~\ref{fig:gfmc} shows the convergence of the calculations of the ground
and Hoyle states in the \AGFMC\ calculations.  Two different sets of
initial states are propagated to $\tau \approx 1.0\,\mbox{MeV}^{-1}$;
they yield consistent results.
The ground-state energy is well reproduced, and the Hoyle state
excitation energy is approximately reproduced 
(see \cite{Epe:2011,Epelbaum:2012qn} for complementary calculations
of the Hoyle state).
The
ground-state form factor of $^{12}$C is also reproduced by these calculations.

Other recent applications of \AGFMC\ include pair momentum distributions
\cite{Wir:2008},
electromagnetic transitions \cite{Per:2007}, and the studies of trapped neutrons
(``drops'')
described in \Sec{sec:neutrondrops}.

\subsubsection{NCSM and \MFDn\ \pagebudget{(Esmond, James, Peter M.) (1 page)}}
\label{sec:ncsm-mfdn}

The measured lifetime of $^{14}$C, $ 5730 \pm 30$ years, is a
valuable chronometer for many practical applications ranging from archeology to
physiology. It is anomalously long compared with lifetimes of other light nuclei
undergoing the same decay process, allowed Gamow-Teller (GT) beta decay. This
lifetime poses a major challenge to theory because traditional realistic
nucleon-nucleon (NN) interactions alone appear insufficient to produce the
effect \cite{aroua}.
Since the transition operator, in leading approximation, depends on the nucleon
spin and charge but not the spatial coordinates, this decay provides a precision
tool to inspect selected features of the initial and final nuclear states. To
convincingly explain this strongly inhibited transition, we need a microscopic
description that introduces all physically relevant 14-nucleon configurations in
the initial and final states and a realistic Hamiltonian.

\begin{figure}[tb]
{\includegraphics[width=.9\columnwidth]{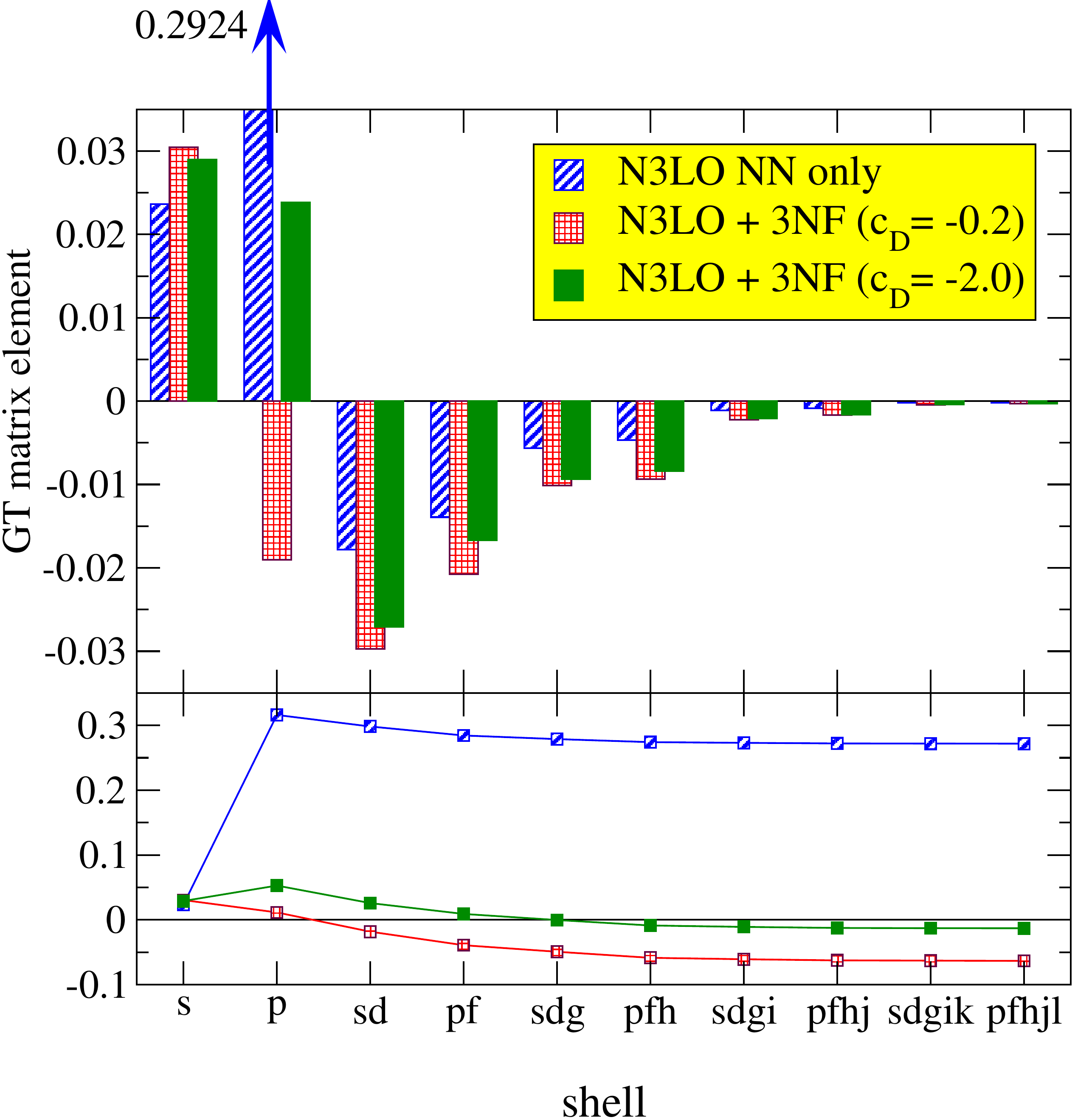}}
\caption{Contributions to the $^{14}$C beta decay matrix element as a
function of the harmonic oscillator 
shell when the nuclear structure is described by a chiral effective field theory
interaction (adopted from \cite{Maris:2011as}). The top panel
displays the contributions with (two right bars of each triplet)
and without (leftmost bar of each triplet) the 3NF  at
$N_{\max} = 8$. 
Contributions  are summed within each shell to yield a total for that shell. The
bottom panel displays the running sum of the GT contributions over the shells.
Note the order-of-magnitude suppression of the $0p$-shell contributions arising
from the 3NFs.}
\label{fig:C14MGT}
\end{figure}

Since the nuclear strong interaction governs the configuration mixing,
the Hamiltonian matrix eigenvalue problem is a very large, sparse 
matrix in the configuration space of 14 nucleons.  We address this
computational challenge with the \MFDn\ code
\cite{SternbergSC08,VarySciDAC09,MarisICCS10,Aktulga_new}.  
Aided by a collaboration with applied mathematicians on scalable
eigensolvers and computational resources on leadership-class machines, 
we are able to solve this
beta decay problem  with sufficient accuracy to resolve the puzzle: the decay is
inhibited by the role of 3-nucleon forces (3NFs) as shown in \Fig{fig:C14MGT}
(see \cite{holtjw2008} for complementary calculations).

We obtained our results on the Jaguar supercomputer (see \Sec{sec:performance}) 
using up to 35,778 hex-core processors (214,668 cores)
and up to 6 hours of elapsed time for each set of low-lying
eigenvalues and
eigenvectors.  The number of nonvanishing matrix elements exceeded
the total memory available and required matrix element recomputation
``on the fly'' for the iterative diagonalization process employing the Lanczos
algorithm.

\begin{figure}[tb]
{\includegraphics[width=\columnwidth]{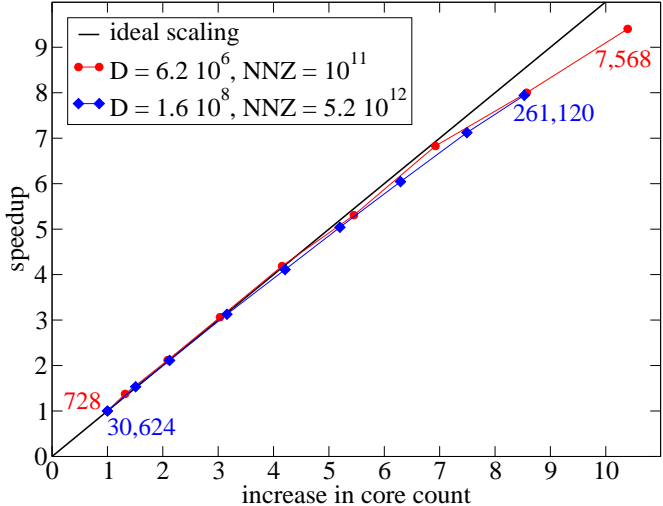}}
\caption{Strong scaling for \MFDn:
speedup for 500 Lanczos iterations (the most time-consuming phase of the code).
Two problems are shown with their dimension 
(D) and number of nonzero matrix elements (NNZ)
in the legend. The smaller is $^7$Li 
(D=6.2 million, NNZ=118 billion),  
and the larger is $^{10}$B 
(D=160 million, NNZ=5.2 trillion).  
The smaller
problem needs at least 1 TB in order to store all nonzero matrix
elements in core and needs, therefore, at least 728
cores to fit the problem in core.  The larger problem needs at least
42 TB, and we used between 30,624 and 261,120 cores for that problem. }
\label{fig:MFDn_Scaling}
\end{figure}

These calculations and many other achievements~\cite{NCSM_review} were
made possible by dramatic improvements to
\MFDn\ capabilities during the UNEDF project~\cite{Maris:2012du}.  
The current scaling performance of {\MFDn} is
demonstrated in \Fig{fig:MFDn_Scaling}. Other recent applications of {\MFDn} 
include the prediction (before experimental confirmation) of the spectroscopy 
of proton-unstable $^{14}$F \cite{Maris:2009bx} and studies of trapped neutrons
(``drops'') with a variety of interactions and other ab initio computational
methods \cite{Maris:2013rgq}.

\subsubsection{NCSM and the resonating group
method\pagebudget{(Sophia, Peter N.)   (1 page)}}
\label{sec:ncsm-rgm}

Weakly bound nuclei, or even unbound exotic nuclei, cannot be understood
by using only
bound-state techniques. Our ab initio many-body approach, no-core
shell model with continuum (NCSMC), focuses on a unified description of both
bound and unbound states. With such an approach, we can simultaneously
investigate structure of nuclei and their reactions.  The method combines
square-integrable harmonic-oscillator basis (i.e., via the NCSM 
\cite{NCSM_review}) accounting for the short- and medium-range
many-nucleon correlations with a continuous basis (i.e., via the NCSM with the
resonating group method
(NCSM/RGM)~\cite{Qua:2008,Qua:2009}) accounting for
long-range correlations between clusters of nucleons. With this technique, we
can predict the ground- and excited-state energies of light nuclei ($p$-shell,
$A{\leq}16$) as well as their electromagnetic moments and transitions, including
weak transitions. Furthermore, we can investigate properties of resonances and
calculate characteristics of binary nuclear reactions (e.g., cross sections,
analyzing powers).

\begin{figure}[tb]
\begin{center}
\includegraphics[width=\linewidth]
{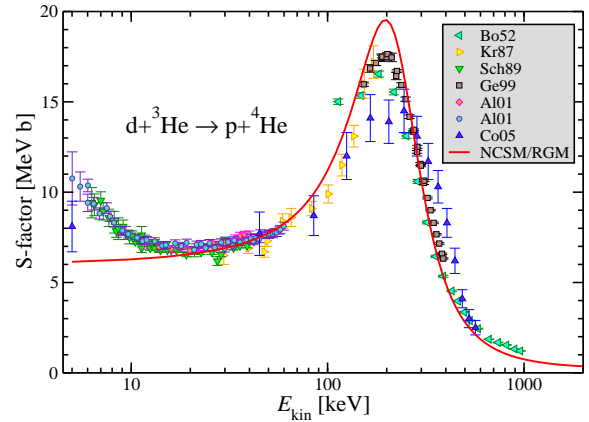}
\end{center}
\caption{\label{fig:d_He3}Experimental results for S-factor of
$^3$He($d$,$p$)$^4$He reaction from beam-target measurements. The full line
represents the  ab initio calculation. No low-energy enhancement is present
in the theoretical results, contrary to the laboratory beam-target data
represented by symbols;
see \cite{NCSMRG2012:navratil} for details.}
\end{figure}

\begin{figure}[htb]
\begin{center}
 \includegraphics[width=\linewidth]
{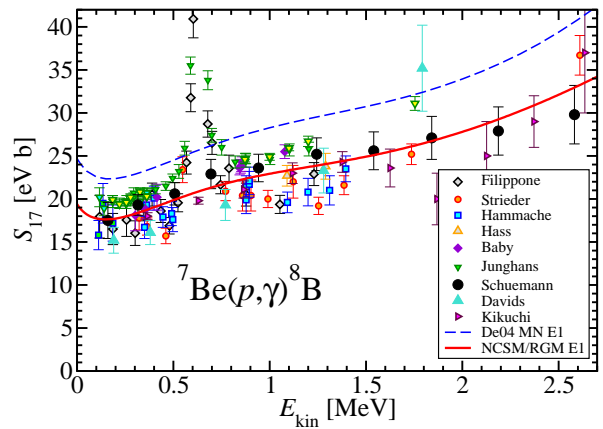}
\end{center}
\caption{\label{fig:p_Be7}Ab initio calculated
$^7$Be($p$,$\gamma$)$^8$B S-factor (solid line) compared with experimental data
and
the calculation used in the latest evaluation (dashed
line); see \cite{Navratil:2011ab} for details.}
\end{figure}

Recent applications of our ab initio techniques include an investigation
of the unbound $^7$He~\cite{PRL110}, calculations of $^3$H($d$,$n$)$^4$He
and $^3$He($d$,$p$)$^4$He fusion~\cite{NCSMRG2012:navratil} (see
\Fig{fig:d_He3}), and calculation of the $^7$Be($p$,$\gamma$)$^8$B radiative
capture~\cite{Navratil:2011ab}, which is important for the standard solar model 
and neutrino physics (see \Fig{fig:p_Be7}). 
We also developed a three-cluster extension of the method to describe 
the Borromean nuclei (e.g., $^6$He and $^{11}$Li).


\subsubsection{Coupled-cluster method \pagebudget{(Hai Ah, Gaute, Thomas) (1
page)}}
\label{sec:cc}
The coupled-cluster
method~\cite{coester1958,coester1960,kuemmel1978,bartlett2007}
exhibits a favorable scaling that grows polynomially with the mass
number of the nucleus and the size of the model space. The UNEDF
collaboration employed an $m$-scheme-based coupled-cluster
code~\cite{dean2004} and an angular-momentum coupled
code~\cite{hagen2010b}. The latter exploits the preservation of
angular momentum and pushed ab initio computation with ``bare''
interactions from chiral effective field theory~\cite{entem2003} to
medium-mass nuclei~\cite{hagen2008}.  Coupled-cluster theory is based
on a similarity-transformed Hamiltonian and employs a nontrivial
vacuum such as the Hartree-Fock state. In practice, one iteratively
solves a large set of nonlinear coupled equations. The exploitation of
rotational invariance considerably reduces the number of degrees of
freedom but comes at the cost of working in a much more complicated
scheme (involving angular momentum algebra) that poses challenges for
a scalable and load-balanced implementation.

\begin{figure}[tb]
\includegraphics[width=0.95\linewidth]{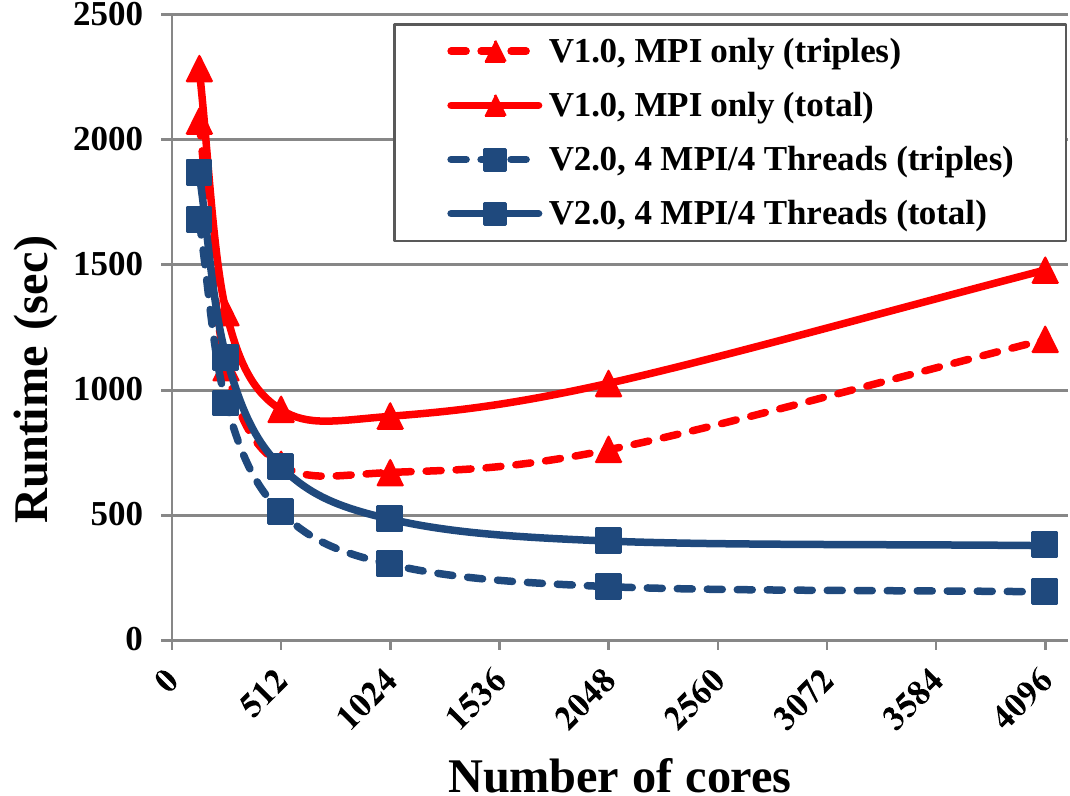}
\caption{Comparison of runtime for $^{40}$Ca in 12 oscillator shells
  using MPI only V1.0 and hybrid MPI/OpenMP V2.0.  Solid lines show
  total runtime; dashed lines show runtime of triples calculation
  only.}
\label{fig:cc_thread}
\end{figure}

During UNEDF, several conceptual advances in physics and computing
were made with the coupled-cluster method. On the physics side, these
include the angular-momentum coupled implementation of the
coupled-cluster method~\cite{hagen2008}, the use of a Gamow basis
for computation of weakly bound nuclei~\cite{hagen2007d,hagen2010a}, 
a practical solution to the center-of-mass problem in nuclear
structure computations~\cite{hagen2009a},  the extension of the
method to nuclei with up to two nucleons outside a closed
subshell~\cite{jansen2011},  the approximation of three-nucleon
forces as in-medium correction to nucleon-nucleon
forces~\cite{holtjw2009,hebeler2010b,hagen2012a}, and the
development of theoretically founded extrapolations in finite
oscillator spaces~\cite{furnstahl2012}. On the computational side,
scaling was improved by a work-balancing
approach~\cite{hagen2009b,hagen2012c} based on MPI and OpenMP such
that the model-space size has increased from ten oscillator shells at
the inception of UNEDF~\cite{hagen2007b} to 20 oscillator shells at
UNEDF's completion~\cite{hagen2012a}.  Figure~\ref{fig:cc_thread}
shows how adding the use of MPI and OpenMP in V2.0 improved the code's
scalability to thousands of cores, beyond a few hundred cores in V1.0
using MPI only, when calculating the small system of $^{40}$Ca in 12
oscillator shells.
We note that the number of single-particle orbitals grows as the third
power with the number of oscillator shells and that the number of
computational cycles -- in the coupled-cluster method with singles and
doubles (CCSD) approximation -- grows as $n_{\rm o}^2n_{\rm u}^4$
(where $n_{\rm o}$ and $n_{\rm u}$ are the numbers of occupied and
unoccupied single-particle states, respectively).  Thus, conceptual and
algorithmic improvements during UNEDF allowed us to solve problems
that na\"ively required an increase of computational cycles by about a
factor 4,000.  The combined efforts culminated in the computation
of neutron-rich isotopes of oxygen~\cite{hagen2012a} and
calcium~\cite{hagen2012b}. 

Doubly magic nuclei are the cornerstones for our understanding of
entire regions of the nuclear chart within the shell model. For this
reason, studies on the evolution of structure in neutron-rich
semi-magic isotopes of oxygen, calcium, nickel, and tin are central to
experimental and theoretical efforts. With
$^{40,48}$Ca being doubly magic nuclei, many studies were aimed at
understanding the structure of the rare isotopes $^{52,54}$Ca and
questions regarding the $N=32,34$ shell
closures~\cite{prisciandaro2001,janssens2002,honma2002,liddick2004,dinca2005}.

\begin{figure}[tb]
\includegraphics[width=0.975\linewidth]{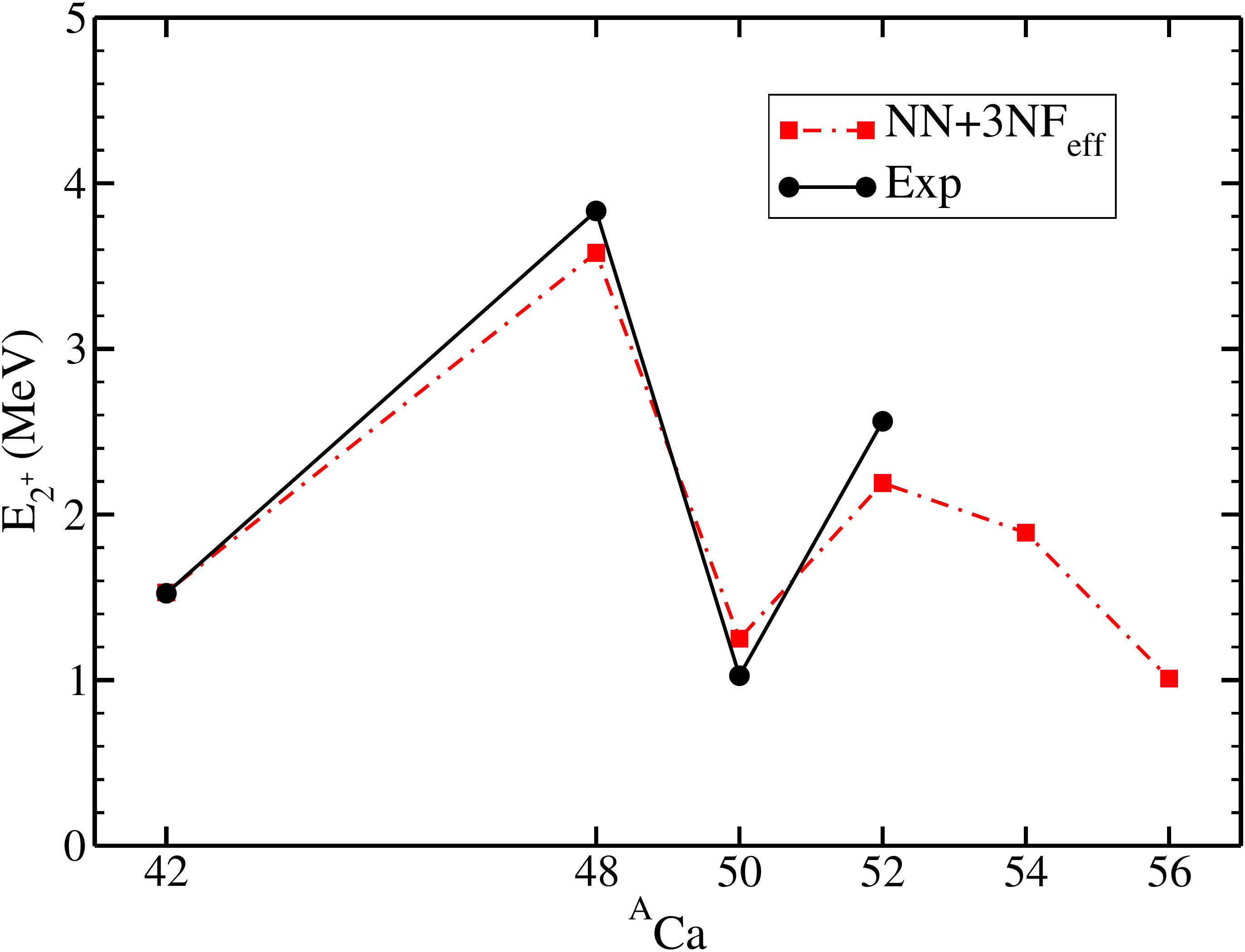}
\caption{Excitation energies of $J^\pi=2^+$ states in Ca isotopes. The
  theoretical results (red squares) agree well with data (black
  circles) and predict a soft subshell closure in $^{54}$Ca. }
\label{fig:cc}
\end{figure}

A first-principles description of rare calcium isotopes is challenging
because it requires the control and understanding of continuum effects
(due to the weak binding) and 3NFs (as often pivotal
contributions arise at next-to-next-to leading order in chiral effective
field theory~\cite{vankolck1994,epelbaum2009,machleidt2011}).
Reference~\cite{hagen2012b} reports coupled-cluster results for
neutron-rich isotopes of calcium that include the effects of the
continuum and 3NFs (see \cite{holt2012} for complementary calculations). It
predicts a soft subshell
closure in the $N=32$ nucleus $^{54}$Ca and an ordering of
single-particle orbitals in neutron-rich calciums that is at variance with
na\"ive shell-model expectations. Figure~\ref{fig:cc} shows the computed
energies of the first excited $J^\pi=2^+$ state in some isotopes of
calcium and compares them with available data. The high excitation
energy in $^{48}$Ca is due to its double magicity, and the somewhat
increased excitation energies in $^{52,54}$Ca suggest that these
nuclei exhibit a softer subshell closure. Where data are available, the
theoretical results agree well with experiment. For $^{54}$Ca, theory
made a prediction that has recently been
verified experimentally \cite{steppenbeck2012}.

\subsection{Configuration interaction \pagebudget{(James, Calvin/Erich, Mihai,
...)
(1 page)}}
\label{sec:ci}

The nuclear shell model has been very effective in describing the physics of 
larger nuclei beyond the current reach of pure ab initio methods; indeed, Eugene
Wigner, Maria Goeppert-Mayer, and 
J.\ Hans D.\ Jensen were awarded the 1963 Noble prize for the fundamental
symmetries and mean field features that underlie the successful
nuclear shell model. 
The shell model for larger nuclei uses the same 
configuration interaction methods 
as the NCSM methods described previously, but with 
more truncated model spaces where not all nucleons are ``active'' and with 
effective interactions tailored for these spaces.

Since there are numerous challenging 
physical applications in nuclear physics that vary across the periodic table,
different CI approaches are needed to efficiently exploit the available
computational resources. 
CI approaches developed or improved within UNEDF
include the following: 
\begin{itemize}
\item No-core shell model in the $m$-scheme basis (\MFDn\  
\cite{SternbergSC08,MarisICCS10,Aktulga_new}; \BIGSTICK\ 
\cite{BIGSTICK,Caurier1,Caurier2}). 
\item No-core shell model in a coupled angular momentum basis (\MFDnJ\  
\cite{AktulgaHPCS11,AktulgaHPSS11}).
\item Shell model with a core in a coupled angular momentum basis ({\NuShellX} 
\cite{nsx-ws,Caurier2}).
\end{itemize}
UNEDF took advantage of common elements in the various CI approaches
to improve the effectiveness of  the nuclear shell model for all nuclei.

These CI  codes utilize an input NN interaction file
and a Coulomb interaction between the protons.  They all work in the
neutron-proton basis (i.e., break isospin) and allow for charge-dependent NN
interactions.
In addition, several of these codes accept 3NFs as
input.  All these codes evaluate the spectra, wavefunctions, and a suite of 
observables for low-lying states of the nucleus.

The implemented algorithms differ considerably among the codes as well as 
support systems for processing the output files generated, such as the
wavefunctions
and one-body density matrices, both static and transition.  Numerous
cross-comparisons
between the codes have been accomplished and their respective accuracies
confirmed.
Eigenenergies are obtained to the accuracy of 1 keV or better. Other observables
are found
to differ at the level of a few percent because of numerical noise in the
wavefunctions.

Except for \MFDnJ\ (which followed {\MFDn}), the codes evolved along independent
paths, which emphasized various strategic physics and technological goals. 
For example, the challenges of addressing heavier nuclei impel working with a
nuclear core; the challenges of working with leadership-class machines versus
local clusters drive some of the algorithmic decisions.  The burden of
communications and memory restrictions 
help resolve the challenge of store-in-memory versus recompute-on-the-fly
strategies that are 
implemented differently in these CI codes.

In light of the need to store large amounts of data for retrieval,
postanalyses, 
and reproducibility, we have developed a prototype database management system.
This prototype records in the database the 
metadata of every run. The data
referenced in the database may include physical observables, one-body
density matrices, and wavefunctions that result from the 
ab initio codes; such data are typically stored on the platforms where
runs are performed.  A user can access this database over the web and
find out whether the runs of interest have already been performed
and where the results may be located.

\subsection{Nuclear density functional theory \pagebudget{(Witek) (1/2 page) }}
\label{sec:dft}

Because of the enormous configuration spaces involved, the properties of complex
heavy nuclei are best described by the superfluid nuclear density 
functional theory \cite{Ben03} -- rooted in the self-consistent
Hartree-Fock-Bogoliubov (HFB), or Bogoliubov-de Gennes,  problem. 
The main ingredient of nuclear DFT is the effective interaction between nucleons
captured by the energy density functional. Since the nuclear many-body
problem involves two kinds of fermions, protons and neutrons,  the EDF depends
on two kinds of densities and currents \cite{[Eng75],Roh10}: isoscalar
(neutron-plus-proton) and  isovector (neutron-minus-proton).  
The coupling constants of the nuclear EDF are usually adjusted  to
selected experimental data and pseudodata obtained from ab initio calculations.
The self-consistent HFB equations allow one to compute  the nuclear ground
state and a set of  quasiparticles that are elementary degrees of freedom of the
system and that can be used to construct better approximations of the excited
states. The HFB equations constitute  a system of coupled 
integro-differential equations that can be written in a matrix form as a 
self-consistent eigenvalue problem, where the dependence of the HFB Hamiltonian 
matrix 
on the eigenvectors (quasiparticle wavefunctions) induces nonlinearities. 

The atomic nucleus is also an open system having unbound states at energies
above the particle emission threshold, and this has implications for the nuclear
DFT. The finiteness of  the HFB potential experienced by a nucleon
implies that the energy spectrum of HFB quasiparticles  contains discrete bound
states, resonances, and nonresonant continuum states \cite{Dob84}. 
The size of the  continuum space may  become intractable, especially for 
complex geometries where self-consistent symmetries are broken.  To this end,
one has to develop methods \cite{Pei11} to treat  HFB resonances and
nonresonant quasiparticle continuum 
without resorting to the explicit computation of all  states.

The application of high-performance computing, modern optimization techniques, 
 and statistical methods has revolutionized nuclear DFT during recent years, 
in terms of both developing new functionals and carrying out advanced
applications. Optimizing the performance of a single HFB run is crucial for
making the EDF
optimization \cite{[Kor10],[Kor12]}  manageable and quickly computing tables of
nuclear observables \cite{Stoitsov03,[Sto09a],[Erl11],erler2012}, in order to
assess theoretical uncertainties. 
  These advances are described in the following
sections.

\subsubsection{DFT solvers \pagebudget{(Stefan/Jason/Nicolas/Witek) (1
pg)}}
\label{sec:dftsolvers}

Solutions of HFB equations can be obtained either by direct numerical
integration on a mesh,
provided proper boundary conditions are imposed on the domain, or by expansion
on a basis. For the latter case, the harmonic oscillator (HO) basis proves
particularly well-adapted to nuclear structure problems, as it offers
analytical, localized solutions with convenient symmetry and separability
features. Although solving the HFB equations for a given nuclear configuration
is
relatively fast on modern computers,  accurate characterization of nuclear
properties often requires simultaneous computations of many different
configurations, from a few dozen (e.g.,  one-quasiparticle
configurations in odd mass nuclei) to a few billion or more in extreme
applications (such as
probing multidimensional potential energy surfaces of heavy nuclei during the
fission process).

The two primary DFT solvers based on HO expansion used by the collaboration are
{\HFBTHO} \cite{Sto05} and
{\HFODD} \cite{Dob97}; see \cite{Sch13} and \cite{Sch11}, respectively, for
their latest releases. Both codes solve the HFB equations for generalized
Skyrme functionals in a deformed HO basis and have been carefully benchmarked
against one another up to the 1 eV level. {\HFBTHO} assumes axial and
time-reversal symmetry of the solutions, making it a very fast program
(execution completes in typically less than 1 minute on a single node). It is
particularly suited for EDF optimization (see \Sec{sec:edfopt})
or large-scale  surveys of nuclear properties  \cite{[Sto09a],[Erl11]}. The
solver {\HFODD} is fully
symmetry-unrestricted: this versatility is necessary for science applications
such as the computation of fission pathways \cite{Sta09} or description of
high-spin states \cite{[Yue12]}.

The new versions of each solver benefited significantly from recent advances in
high-performance
computing and from collaborations with computer scientists in UNEDF. By
expanding the use of tuned {\BLAS} and {\LAPACK} libraries, significant
performance
gains were reported for both codes and enabled new, large-scale studies
\cite{Sch10}. The speed of {\HFBTHO} was further improved by a factor of
2 by
incorporating multithreading; {\HFODD} was turned into a hybrid MPI/OpenMP
program: nuclear configurations are distributed across nodes, while on-node
parallelism is implemented via OpenMP acceleration.

\begin{figure}[tb]
\centerline{\includegraphics[width=0.8\linewidth]{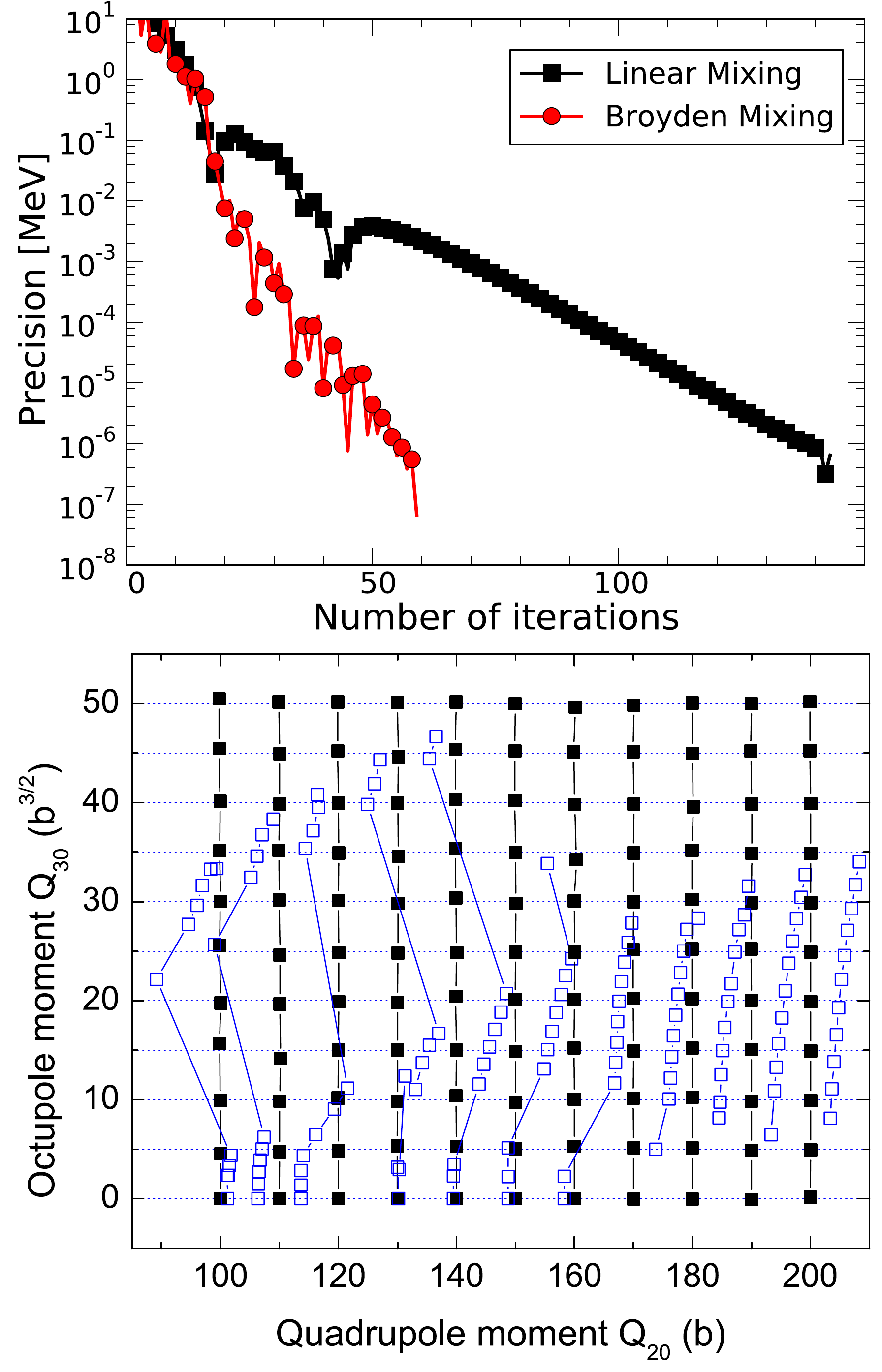}}
\caption{Algorithmic improvements to {\HFODD}. Top: Convergence for a typical
HFB calculation in the ground state of
$^{166}$Dy with {\HFODD} version 2.49t \cite{Sch11}. Using the Broyden
method to iterate the nonlinear HFB equations has provided significant
acceleration compared with traditional linear mixing techniques.
Bottom: Comparison between the augmented Lagrangian method (black squares) and
the standard quadratic penalty method (open squares)  for the constrained  HFB
calculations of the total energy surface  of $^{252}$Fm in a two-dimensional
plane of elongation, $Q_{20}$, and reflection-asymmetry, $Q_{30}$.
(From \cite{Sta10}.)}
\label{fig:dftsolvers}
\end{figure}

Figure~\ref{fig:dftsolvers}
illustrates two
algorithmic improvements to the DFT solver {\HFODD}. The implementation of the
Broyden
method for nonlinear iterative problems \cite{[Bar08]} has reduced
substantially the number of iterations needed to converge the solution in
practical applications. The second example shows the application of the
augmented Lagrangian method (ALM) to fission in $^{252}$Fm \cite{Sta10}.
This method is generally used for  constrained optimization problems; it  allows
precise calculations of multidimensional energy surfaces in the space of
collective coordinates. Indeed, while the standard quadratic penalty method 
often fails to produce a solution at the required values of constrained
variables on a rectangular grid, the ALM performs well in all cases. Both
improvements displayed in \Fig{fig:dftsolvers}
are key to producing realistic large-scale surveys of fission
properties in heavy nuclei on leadership-class computers, where
walltime is limited and expensive.

Another HFB solver developed by UNEDF is {\HFBAX}. It is based on the B-splines
representation of coordinate space and preserves axial symmetry and space
inversion \cite{Pei08}. The solver has been carefully benchmarked with {\HFBTHO}
and used in several applications involving complex geometries, such as fission
\cite{[Pei09]}
and competition between normal superfluidity and Larkin-Ovchinnikov (LOFF)
phases of polarized Fermi gases in extremely elongated traps \cite{Pei10}.
Hybrid parallel programming (MPI+OpenMP) has been implemented in {\HFBAX} to
treat large box sizes that are important  for weakly bound heavy nuclei. 

New generations of DFT solvers will be taking advantage of emerging
architectures, such as GPUs, and new programming paradigms. In particular, the
cost of performing dense linear algebra in both {\HFBTHO} and {\HFODD} can
become prohibitive as the size of the HO basis increases, especially for more
realistic energy functionals involving some form of nonlocality; this
necessitates novel techniques to handle  many-body matrix elements \cite{Par13}.
The massive amount of data generated by large-scale DFT
simulations will also require significant investments in visualization and
data-mining techniques.

\subsubsection{Multiresolution 3D DFT framework \pagebudget{(Fann, Pei, Witek)
(1 page)}}
\label{sec:mr3dhfb}

\begin{figure}[tb]
\begin{center}
\begin{minipage}[t]{.27\linewidth}
\includegraphics[width=.95\linewidth]{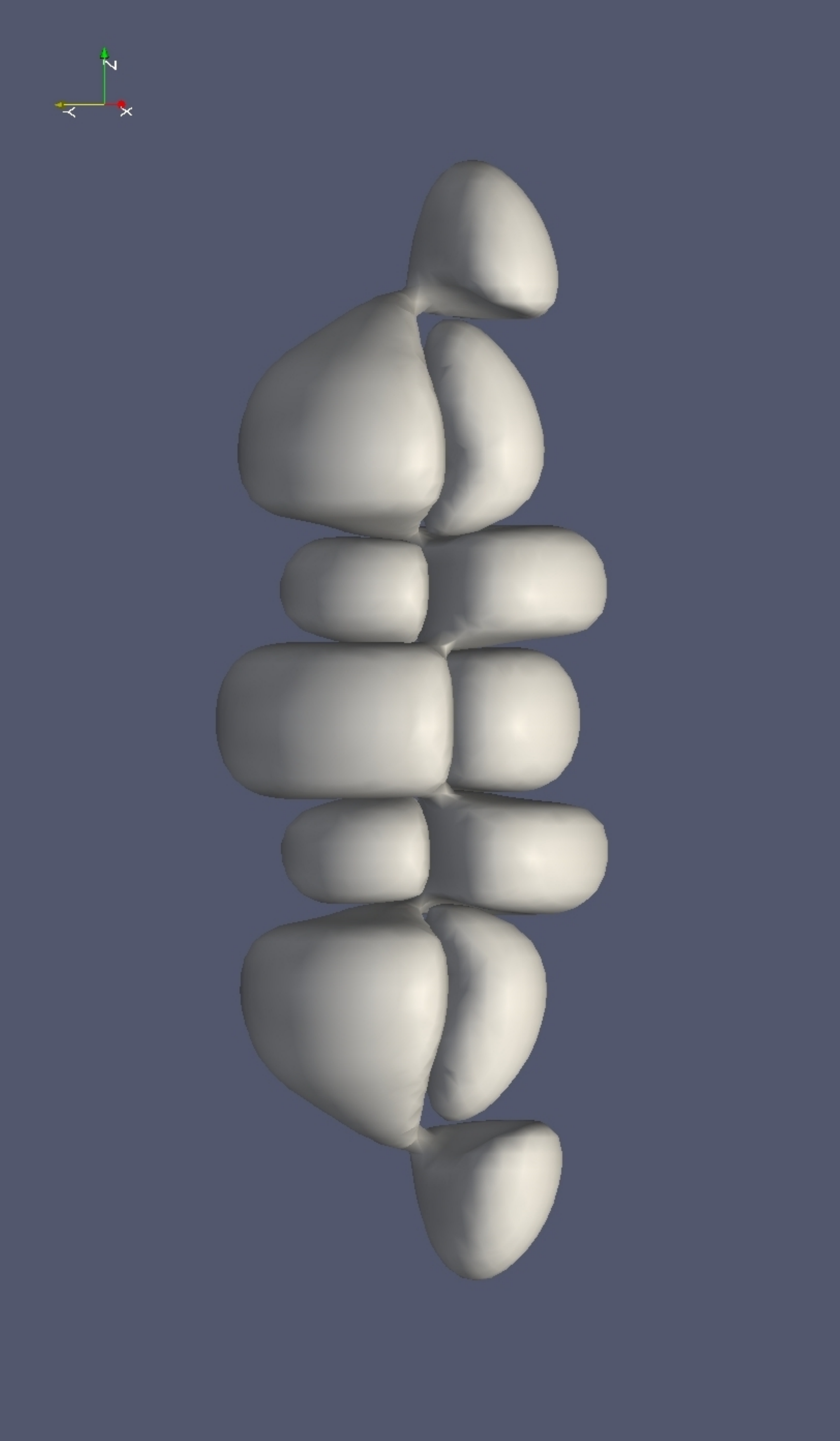} \\
\includegraphics[width=.95\linewidth]{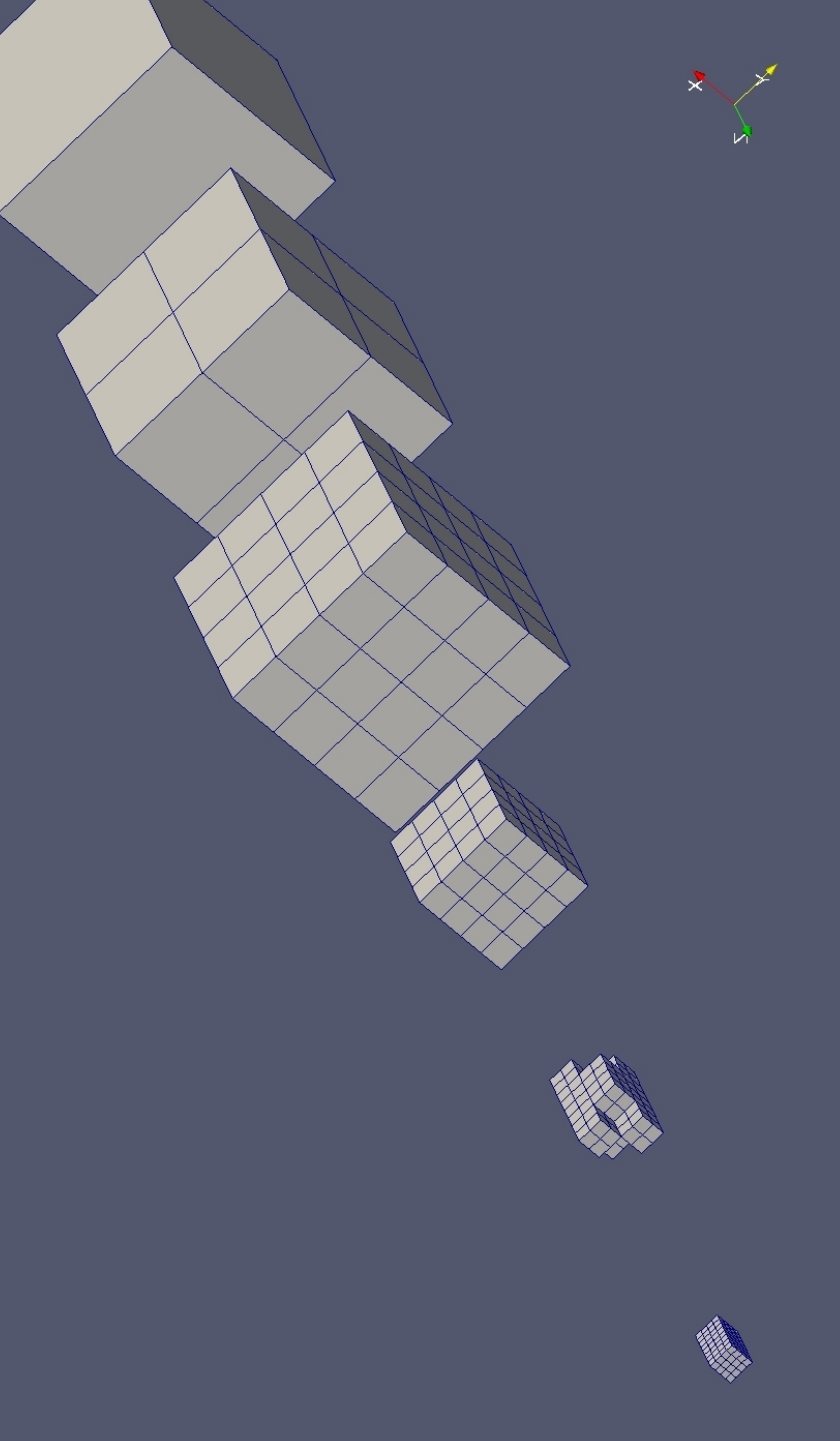}
\end{minipage} \hfill 
\begin{minipage}{.55\linewidth}
 \includegraphics[width=\linewidth]{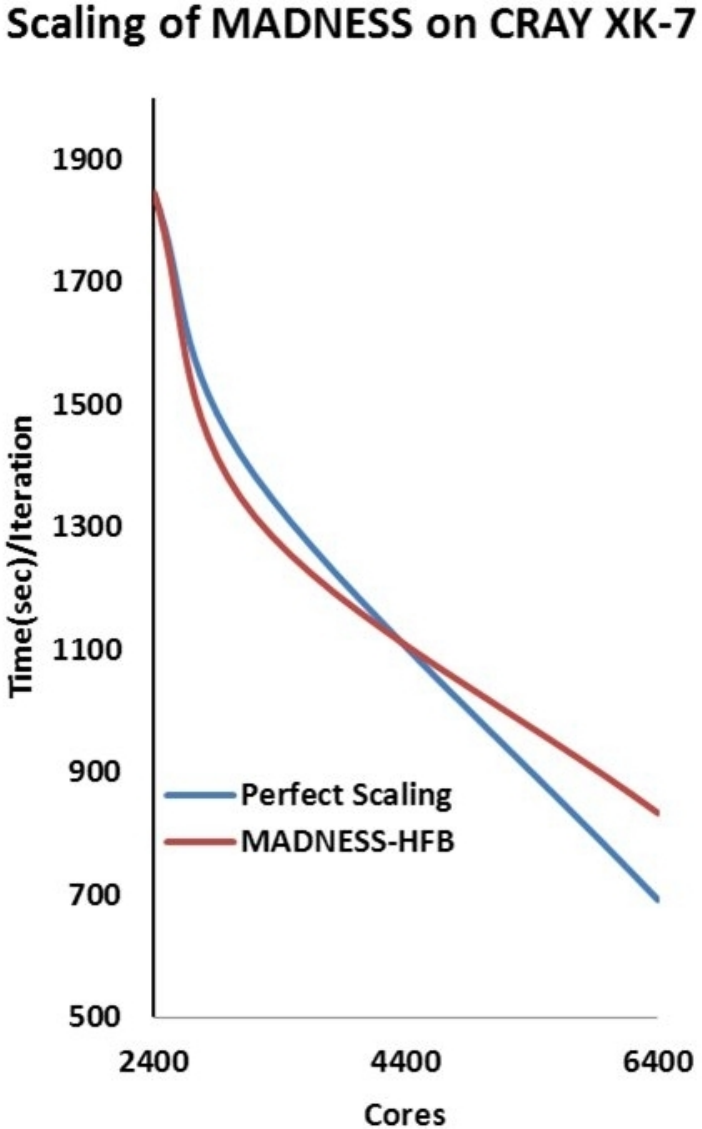}
\end{minipage}
\end{center}
\caption{Quasiparticle wavefunction for a DFT simulation (left, top) and
its
six levels of multiresolution structure (left, bottom).  The refinement
structure is especially
  noticeable at levels 5 and 6.
  Right: The parallel speedup of one iteration of {\HFBMAD}, for solving the
  DFT problem for 1,640 3D quasiparticle wavefunctions with
  over 4.4 billion equations and unknowns;  this simulation was
  performed within a box with a spatial dimension of 120 fermis,
  using 8 multiwavelets, up to level 8+ of refinement, and 
  with a relative precision of $10^{-6}$.
\label{fig:mr3dhfb}
}
\end{figure}

A parallel, adaptive, pseudospectral-based solver,
{\HFBMAD}, has been developed to tackle the fully
symmetry-unrestricted HFB problem for both real and complex wavefunctions in
large
and asymmetric boxes.  The main mathematical and algorithmic advantage
of {\HFBMAD} is its multiscale-multiresolution and sparse
approximation of functions and the application of operators in
coordinate space with guaranteed accuracy but finite precision.
{\HFBMAD} prefers to work with functions and operators with 
pseudo-spectral approximations based on a multiwavelet basis (up to
order 30).  Since the multiwavelets consist of smooth, singular, and
discontinuous functions with spatial locality (compact support), they
are well suited for localized approximation of weak singularities and
discontinuities or regions of high curvature \cite{ABGV02, A-B-C-R:1993,
BR-FA-YA:1997}. 
Gibbs effects are also reduced.  The object-oriented (OO) nature of the
software and template-based programming allow each wavefunction and
each integral or differential operator to have its own boundary
condition and its own sparse pseudospectral expansion.  The usual
boundary conditions (e.g., Dirichlet, Neumann, Robin,
  quasi-periodic, free, and asymptotic conditions) are supported.  Fast
applications of Green's function for
the direct solution of Poisson's equation and the Yukawa scattering
kernel are available \cite{FBHJ04,YFGHB04-2,BCFH07}.  In the
multiwavelet representation, these approximate Green's functions and
their applications are again based on sparse data with guaranteed
precision, in contrast to dense tensors based on the use of some
other basis sets.  Other Green's functions can also be constructed.

If desired, the user can specify solvers and routines from other dense and
sparse linear algebra packages such as {\LAPACK} or {\SCALAPACK}.  For example,
parallel and vectorized adaptive quadrature
permit the construction of the Hamiltonian matrix in the usual manner by 
using the $\ell_2$ norm.  The Hamiltonian can be diagonalized by using
multithreaded {\LAPACK} (or a parallel eigensolver), and the
eigenvectors can be converted back to coefficients for the
multiwavelet representation.
Other capabilities, such as high-order approximation of propagators and
time-stepping required for the solution of time-dependent DFT, are also
available
from applications in time-dependent molecular DFT, as well as from
simulation of attosecond dynamics \cite{PhysRevA.85.033403,JHF2011}.

Underlying this mathematical capability is a parallel runtime system 
that permits the software to scale to hundreds of thousands of
processors and runs on platforms from laptops to leadership-class computers.
The ability to use laptops and workstations is particularly attractive
for model and code development and testing.  In addition, the
embedding of a parser permits the OO-based C++ templated codes
representing operations on the coefficients of each wavefunction to
be executed as parallel tasks.  This parser permits out-of-order, 
 distributed multithread executions with task- and data-dependency analysis. 
This reduces the stalling of execution units
due to data dependencies.  A user-configured and executed parallel
load-balancing method is also available, as is a
parallel checkpoint and restart method.

The 3D {\HFBMAD} has been benchmarked with the spline-based 2D
solver {\HFBAX} \cite{Pei08}, 3D {\HFODD} \cite{Sch11},
and the 1D code {\HFBRAD} \cite{Ben05} for a
variety of problems. Because {\HFBMAD} has no limit on the size of the
computational domain, 
we were able to 
capture quasiparticle wavefunctions with long tails or nonsymmetric potentials
with steep curvatures and
cut-offs to overcome some of the limitations of the other solvers.
The adaptive structure is illustrated in \Fig{fig:mr3dhfb}.

The current {\HFBMAD} approach to the HFB problem is as follows
\cite{FPHJHONSS09}. Let the coefficients of
the wavefunctions in the tensor product multiwavelet representation be
the unknowns.  The user provides an initial relative precision, a set
of initial wavefunctions (e.g., in terms of the HO basis, splines,
etc.), and boundary conditions to start the iterative procedure.  All the
functions, potentials, operators, and expansion lengths are
adaptively represented as needed by the user-defined precision.  A
generalized matrix eigenvalue problem is formed by adaptive
quadrature.  The solution eigenvectors are converted to a sparse
multiwavelet representation for updating the lengths of the expansion
and the coefficients in the potentials, gradients, and other terms
before the next iteration and diagonalization.  The speed and
performance depend on the number of coefficients. Usually, the
simulations begin with a low relative precision, to capture the
low-order terms quickly, before adaptively increasing the order of
approximation and the precision for more accurate results.

\subsubsection{EDF optimization \pagebudget{(Stefan, Jason, Nicolas, Markus,
Witek) (3/4 page)}}
\label{sec:edfopt}

One of the focus areas of UNEDF  was the development of an
optimization protocol for determining the coupling constants of nuclear EDFs. In
particular, the collaboration paid special attention to estimating the errors
associated with such a procedure and exploring the
correlations among the coupling constants. The UNEDF optimization
protocol was established by focusing on the Skyrme energy density. We recall
that, in this framework, the energy of an even-even nucleus in its ground state
is a functional of the one-body density matrix and the pairing tensor. The
Skyrme energy density reads
\begin{eqnarray}
\chi_t(\gras{r})&=& C_t^{\rho\rho} \rho_t^2
  + C_t^{\rho\tau} \rho_t\tau_t +  C_t^{J^2} \bm{J}_t^2
\nonumber \\ &&
  + C_t^{\rho\Delta\rho} \rho_t\Delta\rho_t \
  + C_t^{\rho \nabla J}  \rho_t\bm{\nabla}\cdot\bm{J}_t,
\label{eq:EDF}
\end{eqnarray}
where the isospin index $t$ labels isoscalar ($t$=0) and isovector ($t$=1)
densities, $\rho_{t}$ is the one-body density matrix, and $\tau_{t}$ and
$\bm{J}_{t}$ are derived from $\rho_{t}$ \cite{[Eng75]}. In the pairing channel,
we took a density-dependent pairing energy density with mixed surface
and volume nature, characterized by the two pairing strengths $V_{0}^{(n)}$ and
$V_{0}^{(p)}$ for neutrons and protons, respectively. The set of coupling
constants $C_{t}^{uu'}$, $V_{0}^{(n)}$, and $V_{0}^{(p)}$ are the
parameters $\xb$ to be determined.

\begin{figure}[tb]
\centerline{\includegraphics[width=0.8\linewidth]{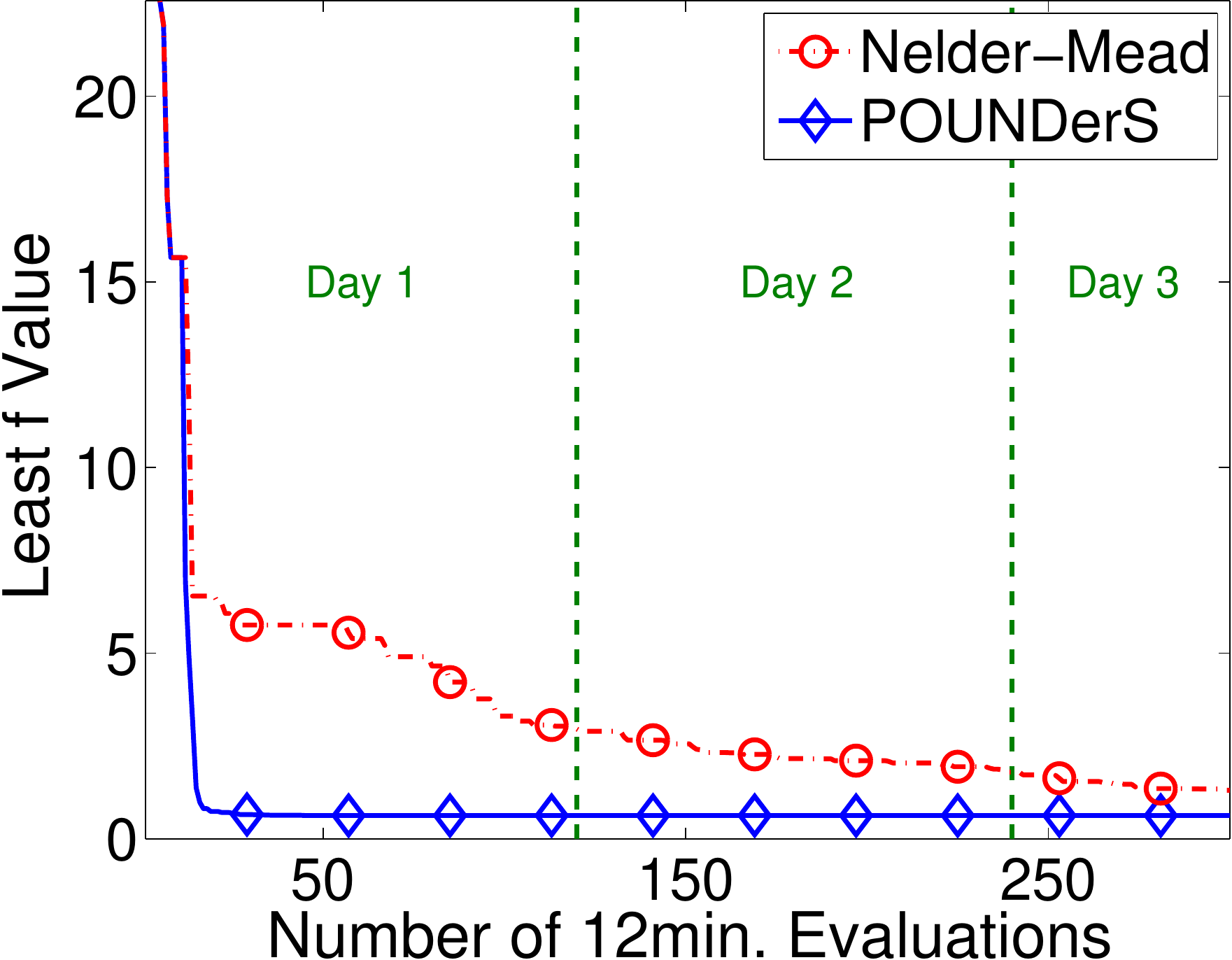}}
\vspace{4pt}
\centerline{\includegraphics[width=0.8\linewidth]{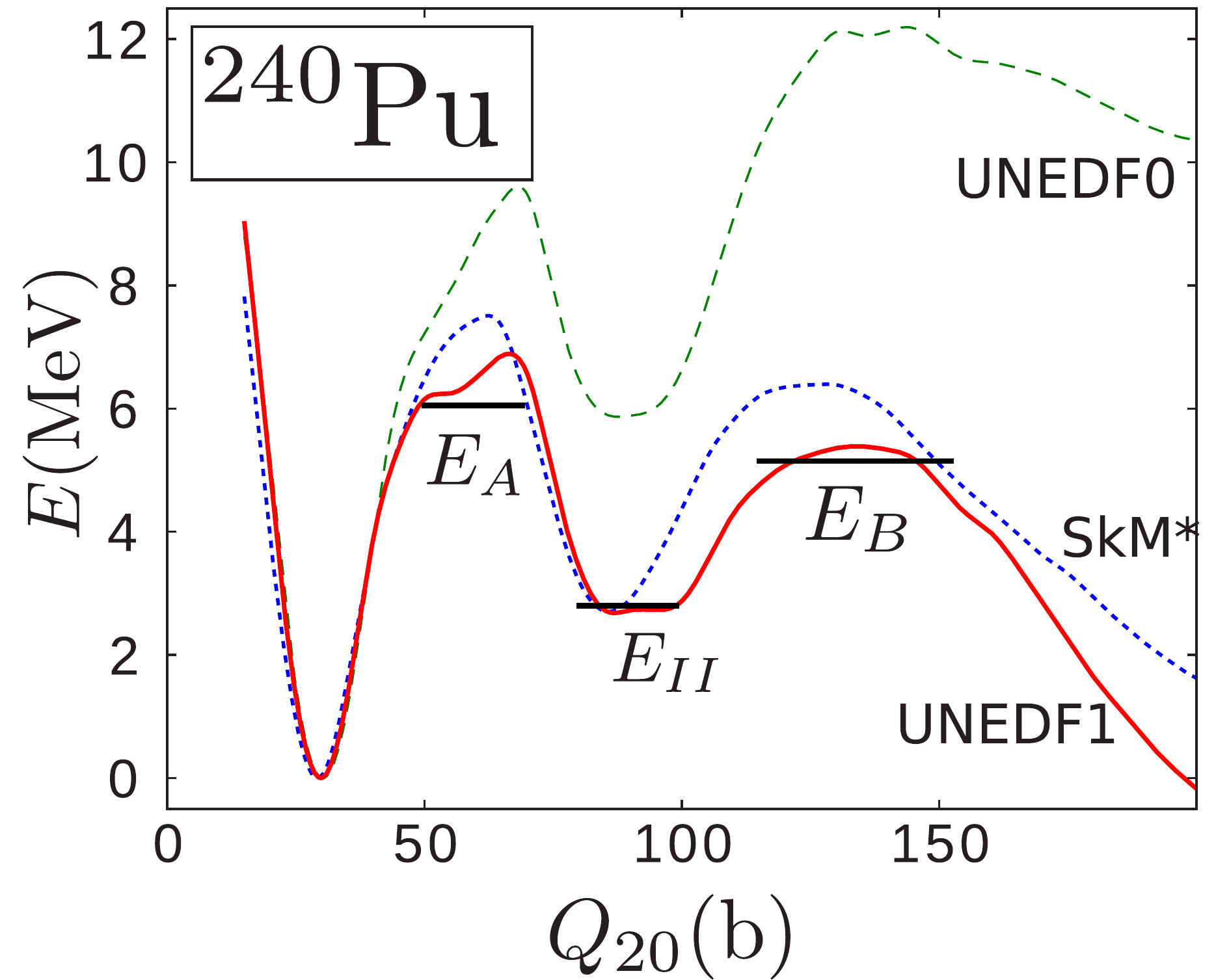}}
\caption{Top: Performance of the \POUNDERS\ algorithm on the minimization of
the $\chi^{2}$ of Eq. (\ref{eq:chi2}) as compared with the standard Nelder-Mead
method. Bottom: Fission pathway for $^{240}$Pu along the mass
quadrupole moment $Q_{20}$ calculated with SkM$^∗$, {\UNEDFZERO},
and {\UNEDFONE} EDFs. The experimental
energy of fission isomer ($E_{II}$) and the inner ($E_A$) and outer ($E_B$) 
barrier heights are indicated \cite{[Kor12]}.}
\label{fig:edfopt}
\end{figure}

The development of fast DFT solvers (see \Sec{sec:dftsolvers}), together
with the availability of leadership-class computers, permitted us for the first
time to set up an optimization protocol at a fully deformed HFB level. Our
first parametrization, {\UNEDFZERO}, was obtained by considering only three
types
of experimental data: nuclear binding energies of both spherical and deformed
nuclei, nuclear charge radii, and odd-even mass differences in selected nuclei
\cite{[Kor10]}. After recognizing that deformation properties needed to be
better constrained \cite{[Nik11]}, a fourth data type, corresponding to 
excitation energies of fission isomers in the actinides, was added. The
resulting parametrization, {\UNEDFONE}, gave a significantly better description
of fission properties \cite{[Kor12]}, see \Fig{fig:edfopt} (bottom). With
the oncoming {\UNEDFTWO}
parametrization, we will expand the optimization data set with single-particle
level splittings. The new data are expected to better constrain the tensor
coupling constants and improve single-particle properties.

Formally, we solve the optimization problem
\begin{equation}
\hspace{-12pt} \min_{\xb}\left \{
\chi^2(\xb)= \sum \limits_{i=1}^{n_d} \left(\frac{ s_{i}(\xb)-d_{i} }{ w_i }
\right)^2 : \xb \in \Omega \subseteq \R^{n_x} \right \}, \,
\label{eq:chi2}
\end{equation}
where $d\in \R^{n_d}$ represents the experimental data,
$w> 0$ represent weights, and the parameters $\xb$ to be determined
are possibly restricted to lie in a domain $\Omega$. This problem is made
difficult because some of the derivatives with respect to the parameters
$\xb$, $\nabla_{\xb} s_i(\xb)$, may be unavailable for some of the
theory simulation observables $s_i$.

Traditional approaches for solving (\ref{eq:chi2}) in the absence of
derivatives typically either estimate these derivatives by finite differencing
or treat $\chi^2$ as a black-box function of $\xb$.
The former approach can be sensitive to the choice of the difference parameter, 
and care must be taken that the expense of the differencing does not grow
unnecessarily as the number of parameters $n_x$ grows. The latter neglects
the structure (in the form of the $n_d$ residuals) inherent to (\ref{eq:chi2}).

In UNEDF, we instead employed a new optimization solver, \POUNDERS, that
exploits the structure in nonlinear least-squares problems and avoids
directly forming computationally expensive derivative approximations.
\POUNDERS\ follows a model-based Newton-like approach, where the first- and
second-order information is inferred by iteratively forming local interpolation
models
for each residual. Figure \ref{fig:edfopt} (top) shows the efficiency of the
solver:
not only does it converge faster than the standard
Nelder-Mead algorithm, but it also gives a more accurate solution. \POUNDERS\ is
available in the open-source Toolkit for Advanced Optimization
(TAO \cite{tao-man}).

\subsubsection{Neutron droplets and DFT \pagebudget{(Stefano, Joe, Markus) (3/4
page)}}
\label{sec:neutrondrops}

The properties of homogeneous and inhomogeneous neutron matter play
a key role in many astrophysical scenarios and in the determination of
the symmetry energy \cite{Gandolfiprc:2012,Tsangprc:2012,Gandolfimnras:2010}. 
The equation of state
of homogeneous neutron matter has been studied in many earlier
investigations (see, e.g., \cite{Gandolfi:2009}).  Since neutron matter
is not self-bound,
inhomogeneous neutron matter has been theoretically investigated by
confining neutrons in external potentials. Although neutron drops cannot
be realized in experimental facilities, they provide a model to study
neutron-rich isotopes \cite{Chang:2004,Gandolfi:2006,Gandolfiepja:2008} and can
bridge
ab initio methods and DFT.
The external potential confining neutrons has been chosen to change
the geometry and density of the system. A Woods-Saxon form produces
saturation, making neutron drops similar to ordinary nuclei. Instead,
a harmonic potential permits one to better control the calculation of
larger systems and to test the approach to the thermodynamic limit.

\begin{figure}[tb]
\begin{center}
\includegraphics[width=.95\linewidth]{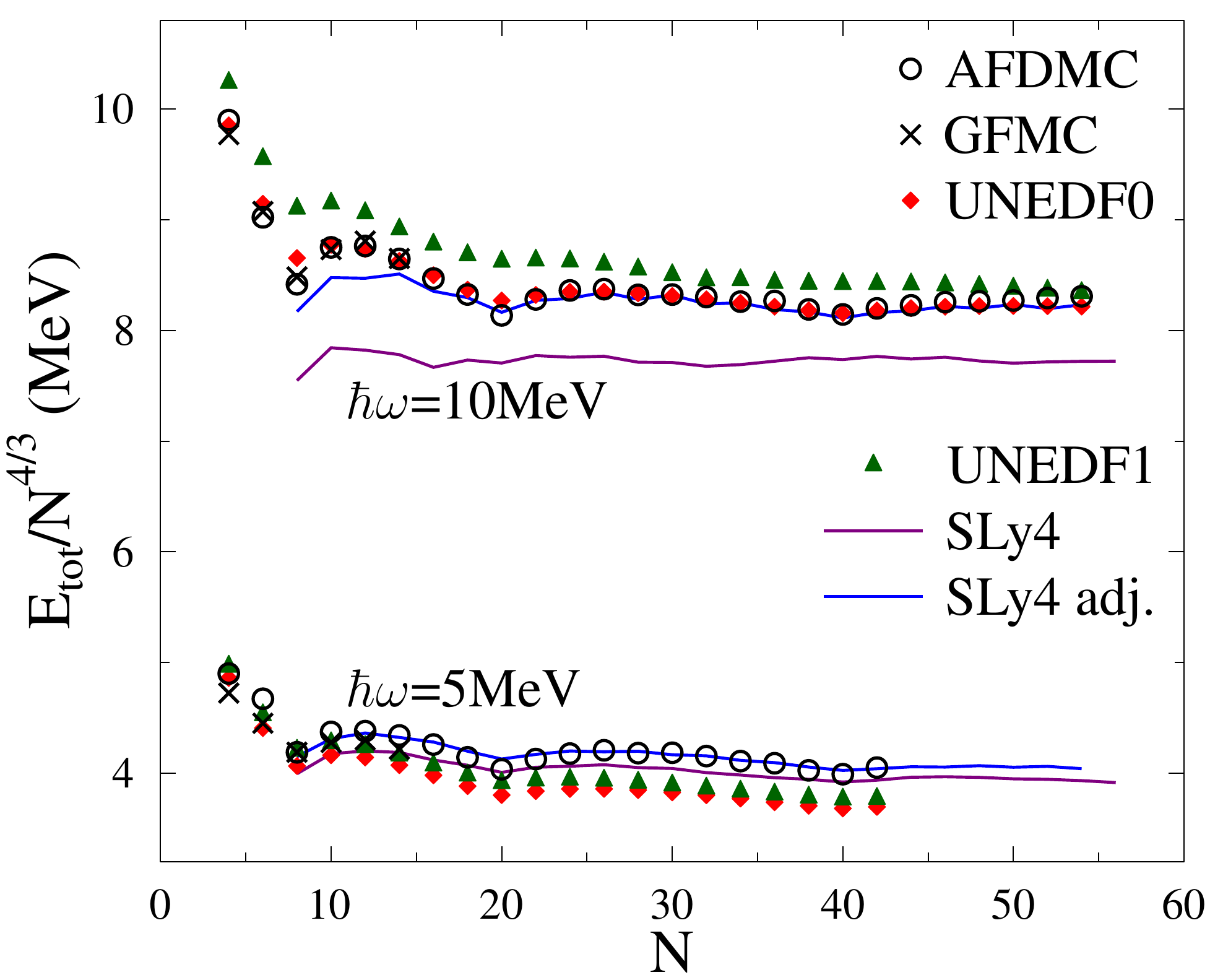}
\end{center}
\caption{Calculated total energies for neutron droplets in
$\hbar\omega=5\,{\rm MeV}$ and $10\,{\rm MeV}$ harmonic potentials as
a function of the neutron number $N$.
The figure shows AFDMC, GFMC, SLy4, and adjusted SLy4 results of
\cite{Gan11b} together with the {\UNEDFZERO} and {\UNEDFONE} results.}
\label{fig:neutrondrops}
\end{figure}

Nuclear EDFs are commonly optimized to reproduce
properties of nuclei close to stability, with close numbers of
protons and neutrons. The use of such functionals to study neutron-rich
nuclei or the neutron star crust requires large 
extrapolations in neutron excess. 
In \cite{Gan11b}, neutron droplets were studied by using QMC methods starting
from a realistic nuclear Hamiltonian that
includes the Argonne AV8' two-body interaction supported by the Urbana IX 
three-body force. This Hamiltonian fits nucleon-nucleon phase shifts,
gives a satisfactory description of light nuclei, and produces an equation
of state of neutron matter that is compatible with recent neutron star 
observations \cite{Steiner:2012}.
The neutron drop's energy calculated by using QMC methods was compared with DFT
calculations. The QMC results
showed that commonly used Skyrme EDFs typically overbind neutron drops 
and that this effect is due mainly to the neutron density gradient term.
The adjustment of the gradient together with the pairing and spin-orbit terms
improves the agreement between ab initio QMC calculations with Skyrme both
for the energy and for neutron densities and radii~\cite{Gan11b}.

These results can be compared with the predictions of {\UNEDFZERO} and
{\UNEDFONE} EDFs.
Figure~\ref{fig:neutrondrops} shows the calculated total energies for neutron
droplets in $\hbar\omega=5\,{\rm MeV}$ and $10\,{\rm MeV}$ harmonic potentials.
The auxiliary field diffusion Monte Carlo (AFDMC) and GFMC QMC results of
\cite{Gan11b}, calculated with the AV8'+UIX
interactions, agree well with the DFT calculations \cite{[Kor12]}. These are 
encouraging,
since neither {\UNEDFZERO} nor {\UNEDFONE}  was optimized to the
pure neutron matter data. Future EDF optimization schemes will use ab initio
results on neutron droplets as pseudodata to improve EDF properties in
 very neutron-rich nuclei.

\subsubsection{
 Ab initio functionals 
\pagebudget{(Scott, Dick,
Markus) (3/4
page)}}
\label{sec:chiraleft}

\begin{figure}[tb]
\begin{center}
\includegraphics[width=0.87\linewidth]{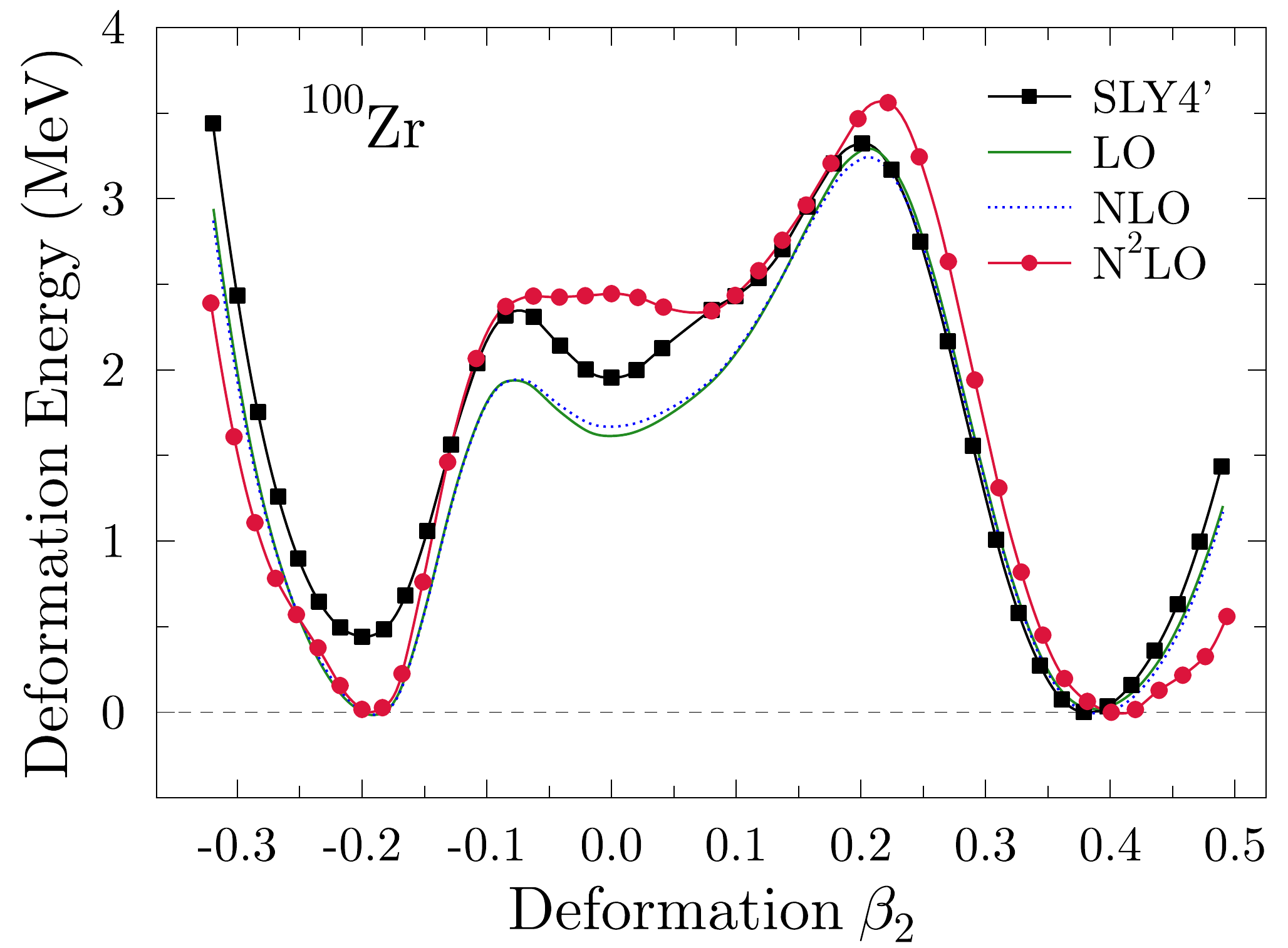}

\vspace*{.2cm}
\includegraphics[width=0.92\linewidth]{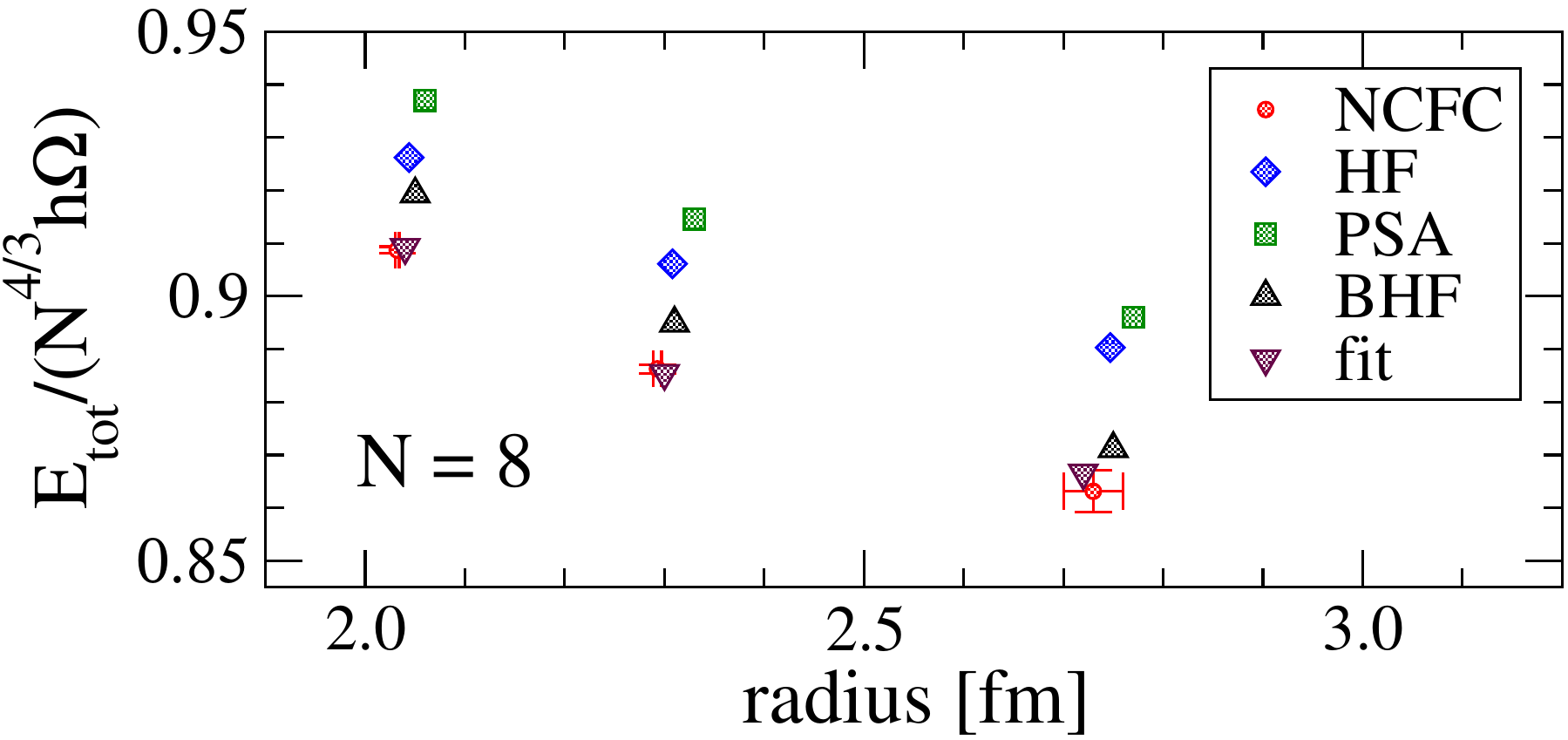}

\end{center}
\caption{
Top: Deformation energy curves for $^{100}$Zr calculated using microscopic
EDFs derived from chiral EFT interactions at different orders \cite{Sto10}.
Bottom: Comparison of microscopic EDF calculations of neutron drops
at increasing levels of approximation with full NCFC calculations starting from
the same Hamiltonian \cite{Bog11}.}
\label{fig:chiraleft}
\end{figure}

In parallel with efforts to improve the optimization of nuclear EDFs with
conventional Skyrme-type terms, UNEDF members sought to construct ab initio
functionals based on microscopic chiral effective field theory
(EFT)~\cite{Drut:2010}.  
A pathway to such functionals was opened with the development of new 
 renormalization group methods, which led to softer 
nuclear Hamiltonians, including three-body forces~\cite{Jurgenson:2009qs}.
These soft interactions dramatically improve
convergence properties in many-body calculations~\cite{Bogner:2009bt},
extending the reach of 
ab initio methods to heavier systems~\cite{jurgenson2011,roth2011a,roth2012}.
At the same time, they make feasible the construction of a microscopically
based EDF using many-body perturbation theory~\cite{Heb10} 
together with improved density matrix expansion (DME)
techniques~\cite{Negele:1972zp,Bogner:2008kj,Geb10}. 
Carrying out this long-term program by individual researchers would be a
formidable 
task, but progress was made possible within UNEDF by teaming up with two of
the physics--CS/AM partnerships described earlier.

An intermediate step toward a fully ab initio EDF was
a new hybrid functional that incorporated long-range
chiral EFT interactions to describe pion-range physics and 
a set of Skyrme-like contact interactions with coupling constants to be fit.
The resulting functional has a much richer set of density dependencies than do 
conventional Skyrme functionals.
These were incorporated in the DFT solvers, and new preoptimization procedures
were developed by the DFT functional group~\cite{Sto10}.
A proof-of-principle test in the top panel of \Fig{fig:chiraleft} shows
deformation
energies in $^{100}$Zr calculated using the DME functional at different
orders in the chiral expansion (LO, NLO, N$^2$LO). 
The deviations from the Skyrme result show
nontrivial effects from the finite-range nature of the underlying NN and
3N interactions~\cite{Sto10}. 
On-going work includes
a rigorous optimization with the procedure outlined in \Sec{sec:edfopt} 
and then detailed evaluations of the predictive power of the DME functional.

In order to directly validate the new DME procedures used in \cite{Sto10}, 
it was necessary to benchmark against
exact results.  The first-ever such calculations 
were made possible by teaming up with the NCSM--\MFDn\ effort
(see \Sec{sec:ncsm-mfdn}) using 
neutron droplets as a controlled theoretical test environment as in
\Sec{sec:neutrondrops}.
The DME functional  was constructed and evaluated for the \emph{same} (model)
Hamiltonian used to
generate exact results from {\MFDn}~\cite{Bog11}
for different numbers of neutrons and varied traps. 
Figure~\ref{fig:chiraleft} (bottom) shows the agreement
between no-core full configuration (NCFC) results and microscopic EDF
calculations at different levels of approximation~\cite{Bog11}, which validates
the optimal strategy
used to construct a microscopically based EDF (the points labeled ``fit''), 
while establishing theoretical error bars.
Further important DME developments 
made by external collaborators in the
FIDIPRO project~\cite{Carlsson:2010da,Dobaczewski:2010qp} will be tested
in future investigations.

\subsection{Beyond DFT \pagebudget{overview: (Engel, Mihai)  (1/2 page)}}
\label{sec:beyonddft}

Static DFT provides excellent tools for investigating nuclear binding energies
and other ground-state properties.  In certain cases, it also can be used to
treat dynamical processes.  The path to scission during fission, for
example, sometimes can be predicted accurately by static DFT.  A reliable
description
of excitation/decay and reactions, however, usually requires methods that go
beyond static DFT.  Since an ab initio treatment of the nuclear time-evolution
is difficult, we employ extensions of DFT and related ideas.  The
simplest extension, the quasiparticle random phase approximation (QRPA), can
be viewed as an adiabatic approximation to the linear response in
time-dependent DFT.  It provides the entire spectrum of excitations with the
same EDF used in static DFT.  The adiabatic
approximation is, of course, severe (as are the approximations in the density
functional itself) but can be applied in any nucleus and folded with reaction
theory.  DFT-based QRPA and its applications to nuclear excitation and
reactions are discussed in \Sec{sec:qrpa}.

DFT-based methods that go beyond the adiabatic approximation are also now in
use.  One can exploit the relatively simple dynamics of Fermi gas systems to
construct an approximate time-dependent extension of DFT, the time-dependent
superfluid local density approximation (TDSLDA).  The approximation and related
computational techniques can be applied to such classic problems as
photoabsorption but also to other time-dependent processes that go beyond
linear response.  The TDSLDA and its applications are discussed in
\Sec{sec:tddft}.

We also need efficient methods to accurately compute average properties of
excited states, such as spin- and parity-dependent level densities, which 
suffice to treat reactions that proceed primarily through a compound nucleus.
Obtaining these densities through a direct diagonalization of the nuclear
Hamiltonian and a subsequent level counting is not efficient, but several
techniques based on statistical spectroscopy can be used instead. However,
even statistical spectroscopy poses computational challenges that demand 
high-performance computational techniques and resources.
Some advances in computational spectroscopy, leading to the first accurate
calculation of densities of levels with unnatural parity, are described in
\Sec{sec:levelden}.

\subsubsection{QRPA and reactions \pagebudget{(Engel, Terasaki, Thompson) (1
page)}}
\label{sec:qrpa}

Members of the UNEDF 
collaboration developed and exploited both an extremely accurate spherical
Skyrme QRPA code \cite{Ter05} and an equally accurate, though computationally
much more intensive, deformed (axially symmetric) Skyrme QRPA code
\cite{Ter10}.  The latter, which can treat both spherical and deformed nuclei,
is at the forefront of the modern QRPA.  Other groups have developed their own
versions of the deformed Skyrme, Gogny, or relativistic QRPA
\cite{[Yosh09],[Ebat10],[Pena09],[Peru11],[Losa10]};
most of these have some disadvantages compared with ours (e.g., a lack of full
self-consistency, oscillator bases that don't capture continuum physics,
etc.) but also the occasional advantage (e.g., full continuum wavefunctions
rather than the approximate representation of the continuum we describe below).

Both our spherical and deformed codes diagonalize the traditional QRPA A-B
matrix \cite{Rin80}, constructed from single-quasiparticle states in the
canonical basis \cite{Rin80} in a large box (typically 20 fm in each
coordinate), so that continuum states are taken into account in discretized
form. Both codes work with arbitrary Skyrme density functionals plus delta
pairing, include all rearrangement terms, and break neither parity nor
time-reversal symmetries.  Both output transition amplitudes to the entire
spectrum of excited states.  

The two codes have some differences as well.  The spherical code gets its
single-quasiparticle wavefunctions, represented on an equidistant mesh, from
an HFB program called {\HFBMARIO}, which derives from the code
{\HFBRAD} \cite{Dob84}. The deformed code takes its wavefunctions from
the Vanderbilt HFB program \cite{Bla05}, which uses B-splines to represent wave
functions.  Each QRPA code represents those wavefunctions in the same manner
as the HFB programs it relies on.

Both QRPA codes have been tested in many ways, including against one another.
With the spherical code, we calculated energy-weighted sums in
Ca, Ni, and Sn isotopes from the proton drip line through the neutron drip line
for $J^\pi=0^+,$ $1^-$, and $2^+$ multipoles with Skyrme parameter sets
SkM$^\ast$ and SLy4 and found excellent agreement with analytical values
\cite{Ter05}.  Spurious states in the $J^\pi=0^+$, $1^+$, and $1^-$ channels
are well separated from physical states in both codes, though the spherical one
performs a bit better because it can include all combinations of HFB
two-quasiparticle states in the QRPA basis without making the calculation
intractable. 

The collaboration used the spherical QRPA to study systematics of $2^+$ states
across the table of isotopes and for microscopic calculations of reaction
rates; they used the deformed version for a more limited study of $2^+$ states
and giant resonances in rare-earth nuclei \cite{Ter11}.  

\begin{figure}[tb]
\begin{center}
\includegraphics[width=0.97\linewidth]{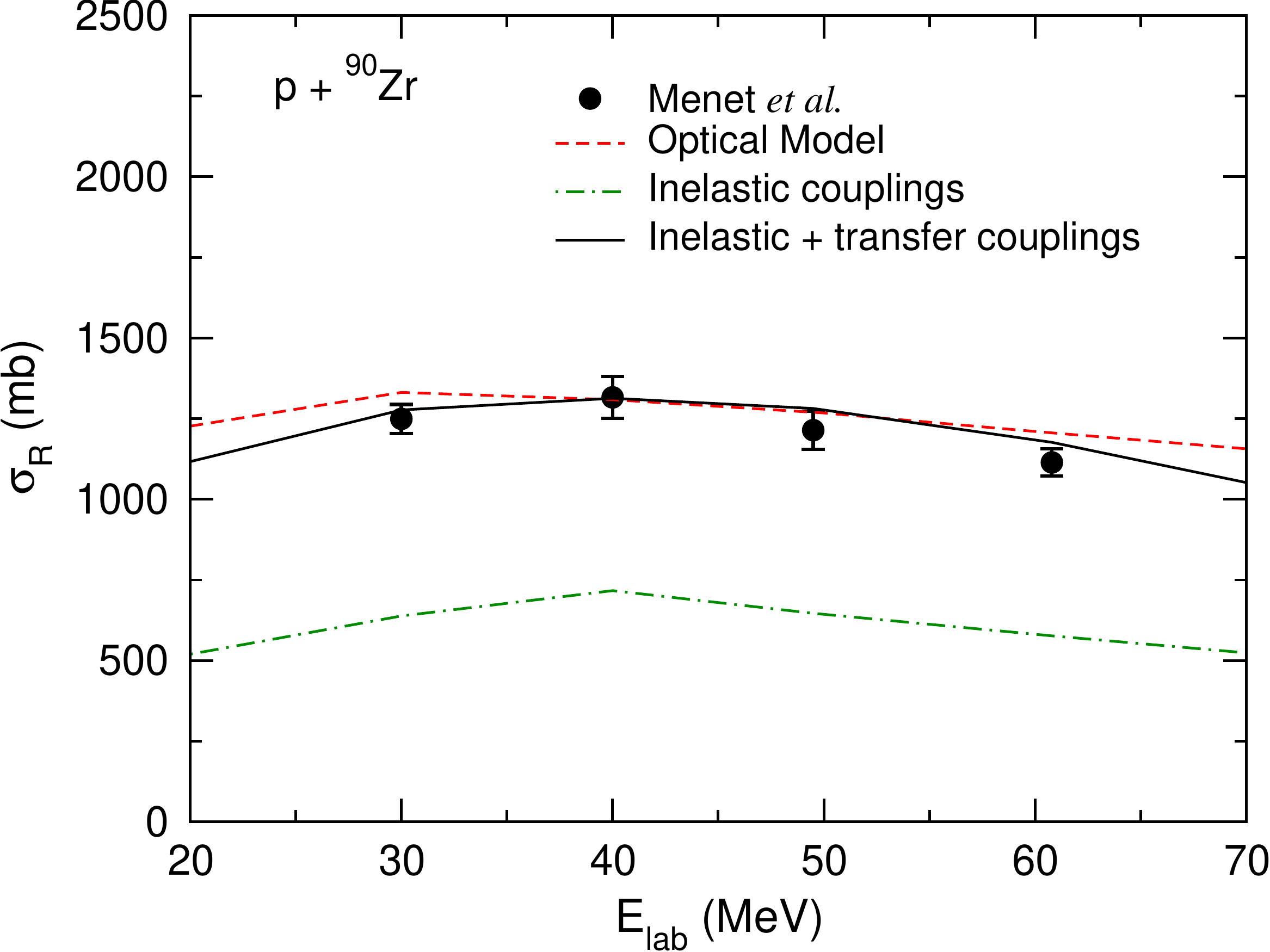}
\end{center}
\caption{Reaction cross section as a function of incident energy for $p$ +
$^{90}$Zr. The results are shown for couplings to the inelastic states
(dash-dotted line) and to the inelastic and transfer channels with
nonorthogonality corrections (solid line). The Koning-Delaroche
\cite{Koning2003NPA713} optical model calculations are also shown (dashed
line). 
(Data from \cite{PhysRevC.4.1114}.)} \label{fig:pzr90abs}
\end{figure}

The collaboration also used transition densities from the spherical QRPA to
calculate nucleon-nucleus scattering.  The transition amplitudes produced by
our spherical matrix QRPA, when combined with single-particle wave functions,
yield radial transition densities.  These can in turn be folded with the
interaction between the projectile and the nuclear constituents (i.e., the
nucleon-nucleon interaction) to produce transition potentials that excite
target states.  References
\cite{PhysRevLett.105.202502,PhysRevC.84.064609,JourPhys.312.082033} report the
development of a code to fold the densities for all QRPA states below 30 MeV
with a Gaussian-shaped nucleon-nucleon potential.  The result is a microscopic
coupled-channels calculation that successfully produces angular distributions
and inelastic cross sections for nucleon-induced reactions---quantities that
can be compared directly with scattering data---at scattering energies between
10 and 70 MeV.  To satisfactorily describe observed absorption, we had to
explicitly couple also to all one-nucleon pickup channels leading to
intermediate deuteron formation. Figure~\ref{fig:pzr90abs} illustrates the
effect of such couplings on nucleon-induced absorption cross sections.  The
direct connection between the calculated cross sections and the nuclear
structure ingredients makes this kind of reaction calculation a good test of
the structure model.  

The collaboration also took significant steps to develop a much more
efficient implementation of the QRPA.  The finite amplitude method 
\cite{[Avog11],[Avog13]} allows one to effectively take the derivatives of mean
fields
that enter the QRPA equations numerically, through relatively straightforward
modifications to the mean-field codes themselves.  A simple iterative procedure
then solves the equations.  Our initial application, to monopole resonances in
the deformed nucleus $^{240}$Pu \cite{[Stoi11]}, consumes a small fraction of
the time our matrix QRPA implementation would use (see
\cite{Toivanen10,Carlsson12} for complementary work based on iterative
Arnoldi diagonalization).

\subsubsection{Time-dependent DFT for superfluid systems
\pagebudget{(Aurel, Kenny) (1 page)}}
\label{sec:tddft}
The application of DFT to nuclear physics requires two nontrivial elements: the
ability to describe both superfluidity and time-dependent phenomena.  
In order to avoid the nonlocal character of the
DFT extension to superfluid systems,  the superfluid local density approximation
(SLDA) and its time-dependent extension TDSLDA have been developed 
\cite{BY:2002fk,Bulgac:2002uq,Bulgac:2007a,BF:2008,Bulgac:2009b,Bulgac:2011,
Bulgac:2011b,Bulgac:2011c,Stetcu:2011,Bulgac:2013b}.

\begin{figure}[tb]
\includegraphics[width=0.32\linewidth]{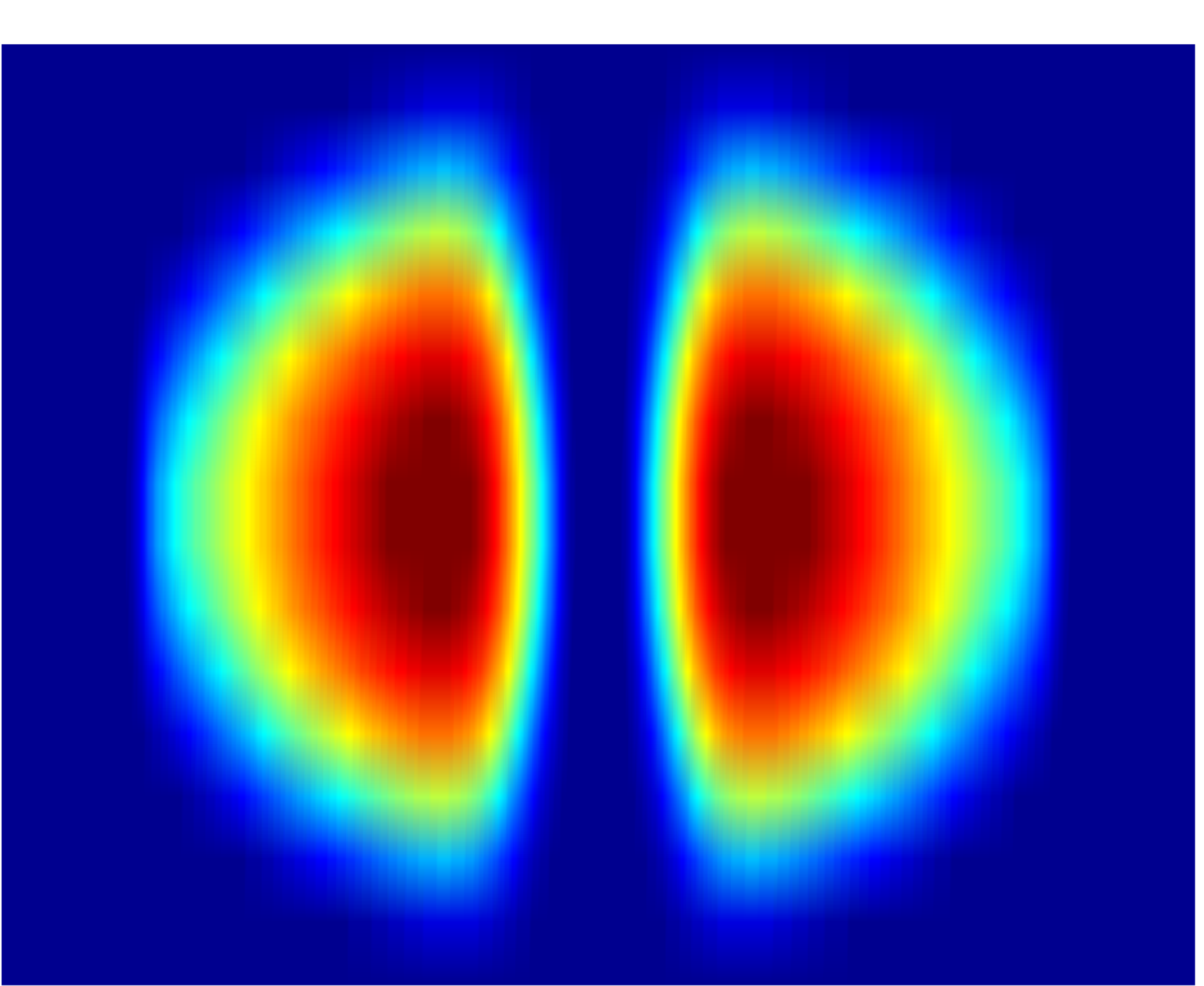} 
\includegraphics[width=0.32\linewidth]{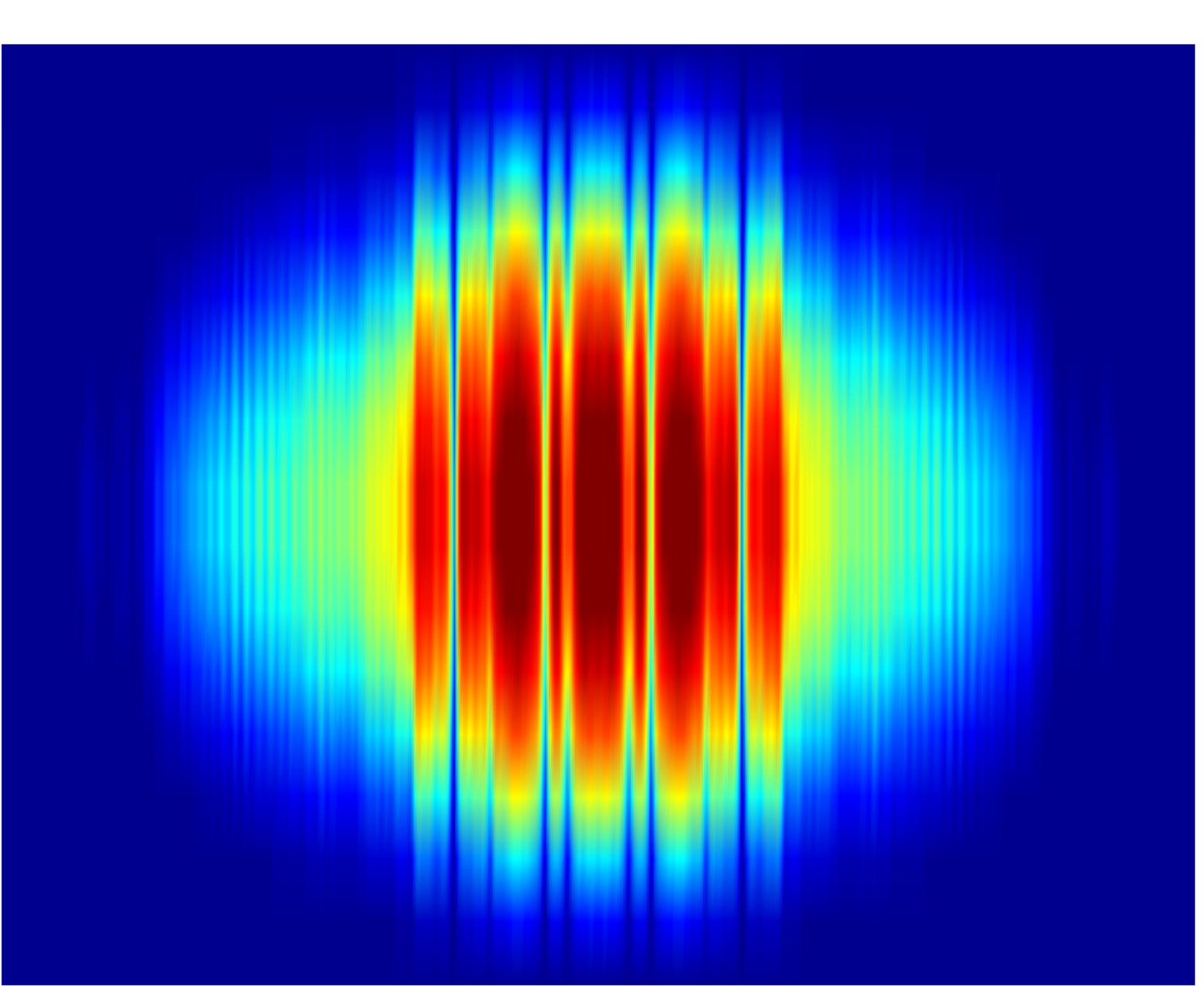} 
\includegraphics[width=0.32\linewidth]{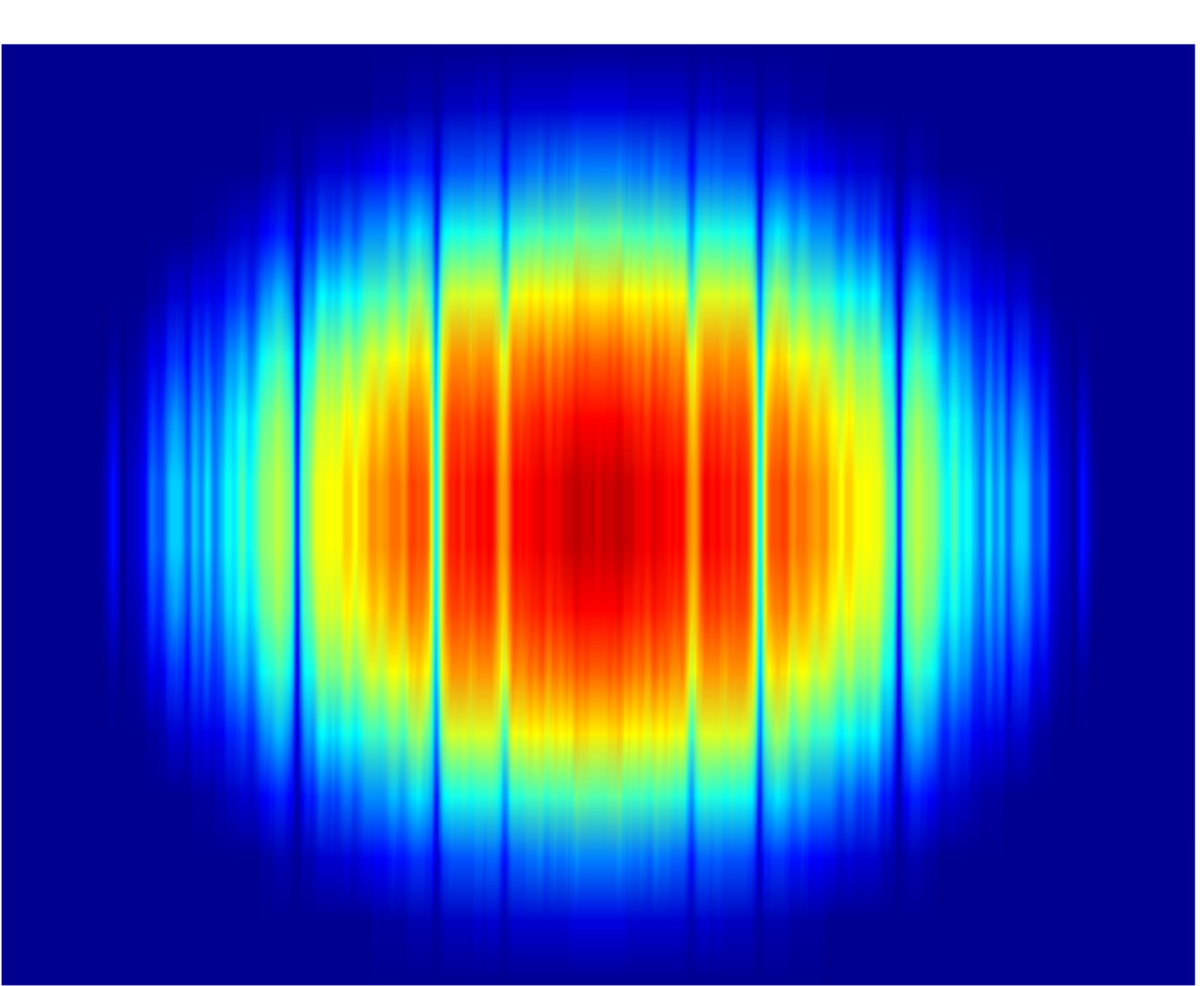}
\caption{Three consecutive frames of the head-on collision of two fermion clouds
of $\approx$750 particles in which quantum shock waves and domain
walls/solitons (topological excitations) are formed \cite{Bulgac:2011c}. The
$x$- and $y$-directions have an aspect ratio of $\approx$30.}
\label{fig:frames}
\end{figure}

SLDA and TDSLDA  have been applied to a large number of
fermionic systems and phenomena: vortex structure in neutron matter and cold
atomic systems, generation and dynamics of quantized vortices and their crossing
and reconnection, excitation of the Anderson-Higgs modes, the LOFF
phase, quantum shock waves and excitation of
domain walls, one- and two-nucleon separation energies, giant dipole resonance
in superfluid triaxial nuclei, and complex collisions. In \Fig{fig:frames}, we
illustrate the case of a head-on collision of two superfluid fermion clouds,
which was studied experimentally. Both SLDA and TDSLDA are derived by using
appropriately determined EDFs with QMC input for homogeneous systems and
validating the predictions on
independent  QMC calculations of inhomogeneous systems in the well-studied case
of a unitary Fermi gas; see \cite{Bulgac:2007a,BF:2008,Bulgac:2011} for
details. The form of the EDF for a unitary Fermi gas is largely determined by
dimensional arguments; translational, rotational symmetry, and parity; gauge and
Galilean covariance (which specifies the dependence on current densities); and
renormalizability of the TDSLDA formalism.

For nuclear systems we lack ab initio results of the same quality and rely
on a more phenomenological approach, but with significant microscopic input. The
nuclear EDFs  should satisfy the usual symmetries \cite{Roh10} and the
consistency
with the best available ab initio results. 

\begin{figure}
\begin{center}
\includegraphics[width=0.75\linewidth]{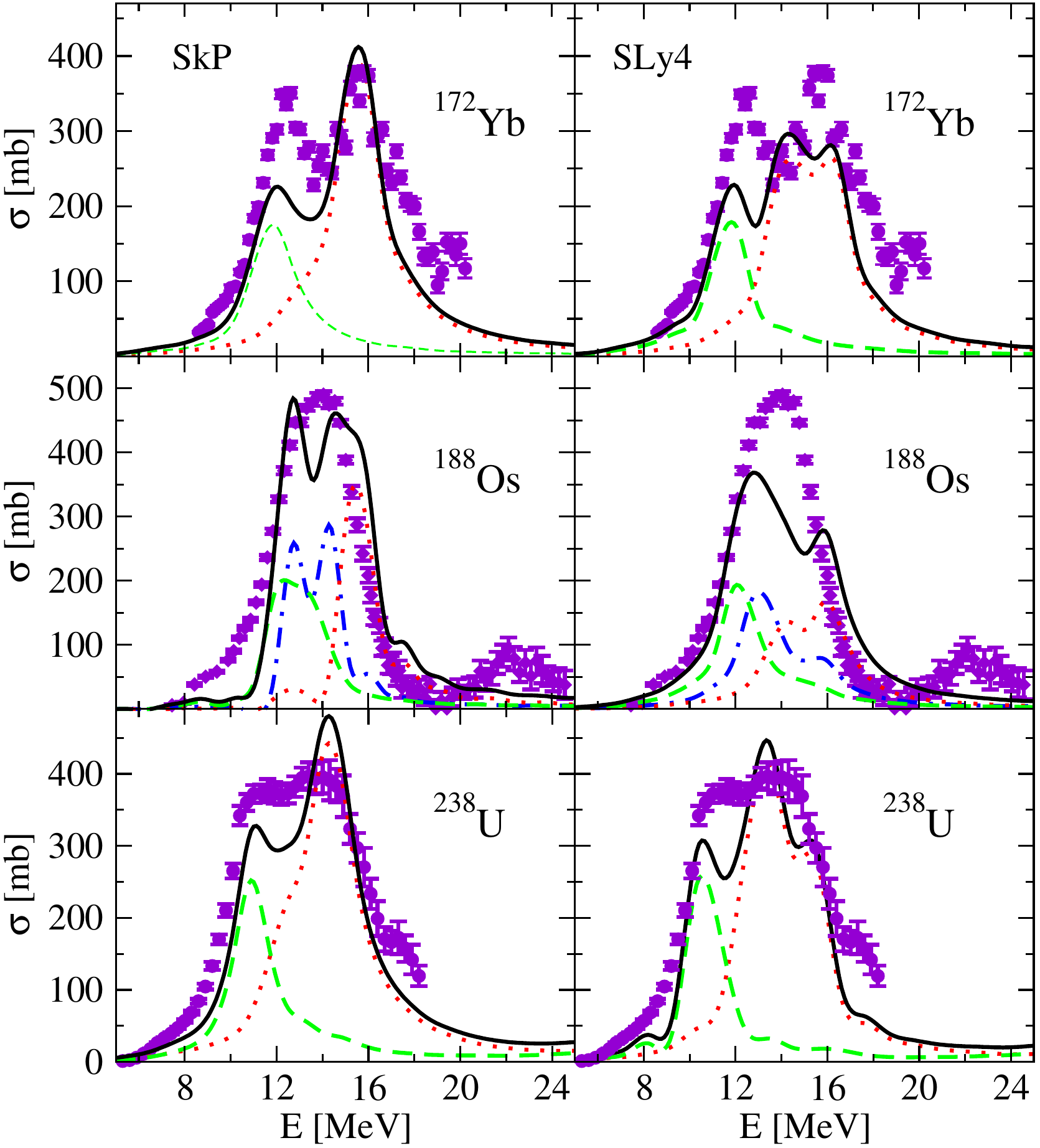}
\end{center}
\caption{Photoabsorption cross section (solid black line) calculated within
TDSLDA using two Skyrme force parameterizations for three deformed open-shell
nuclei and the experimental $(\gamma, n)$ cross sections (solid purple circles
with error bars); see \cite{Stetcu:2011} for details. With dashed (green),
dotted (red), and dot-dashed (blue) lines, we display the contribution to the
cross section arising from exciting the corresponding nucleus along various
symmetry axes.}
\label{fig:gdr}  
\end{figure}

The numerical implementation of
the SLDA and TDSLDA equations leads to hundreds of thousands of coupled
nonlinear 3D time-dependent PDEs, which are solved by 
using the discrete variable representation approach
\cite{Bulgac:2008,Bulgac:2013} on desktops
\cite{Bulgac:2007a,BF:2008,Bulgac:2009b,Bulgac:2011} and---as a result of UNEDF
collaborations with computer scientists---leadership-class supercomputers 
\cite{Bulgac:2011,Bulgac:2011b,Bulgac:2011c,Stetcu:2011,Bulgac:2013b}. In
\Fig{fig:gdr} we illustrate the first calculation of the photoexcitation of a
triaxial superfluid nucleus performed within TDSLDA ($^{188}$Os) and two 
other axially deformed nuclei, as well as a comparison with the absolute
experimental data  (without any fitting parameters).  The determination of the
ground-state properties of these nuclei and their subsequent time-evolution
required full
diagonalizations of Hermitian matrices of sizes up to $5\cdot 10^5\times 5\cdot
10^5$ and
solutions of $5\cdot 10^5 $ coupled time-dependent 3D PDEs.   Further studies
of
excitation of medium- and heavy-mass nuclei with $\gamma$-rays, neutrons,
relativistic heavy ions, and induced nuclear fission  are the next steps. 

\subsubsection{Level densities \pagebudget{(Mihai) (1/2 page)}}
\label{sec:levelden}

The properties of the excited states of nuclei are key to reliably describing 
reactions and decays. One important type of reaction mechanism is the
compound nuclear reaction, which can be described with the statistical model of 
Hauser and Feshbach \cite{Hauser52}. The important ingredient entering the
Hauser-Feshbach theory
is the spin- and parity-dependent nuclear level density (NLD). Experimental
information about
NLD is limited for stable nuclei and not available for radioactive nuclides of
interest for
nuclear astrophysics. Therefore, a large effort is underway to accurately
calculate NLD, and an 
interacting shell model approach would be the best model taking into account the
relevant
many body correlations beyond DFT.
A direct approach by  direct CI diagonalization and level counting is not
feasible because of the exponential increase in CI dimensions.
We recently proposed \cite{scott10}
an approach to calculate shell model spin- and parity-dependent NLD using
methods of statistical spectroscopy. In addition, we showed \cite{horoi07} how
one can improve this approach
to calculate the unnatural parity NLD by removing the contribution due to the
spurious center-of-mass excitations.
The associated algorithms were implemented in a high-performance computer code,
{\JMOMENTS}, 
\cite{sen'kov10,sen'kov11,sen'kov13}, which runs on massively parallel
computers and scales well up to 10,000 processors \cite{sen'kov10,sen'kov13}.

\begin{figure}[tb]
\begin{center}
\includegraphics[width=0.90\linewidth]{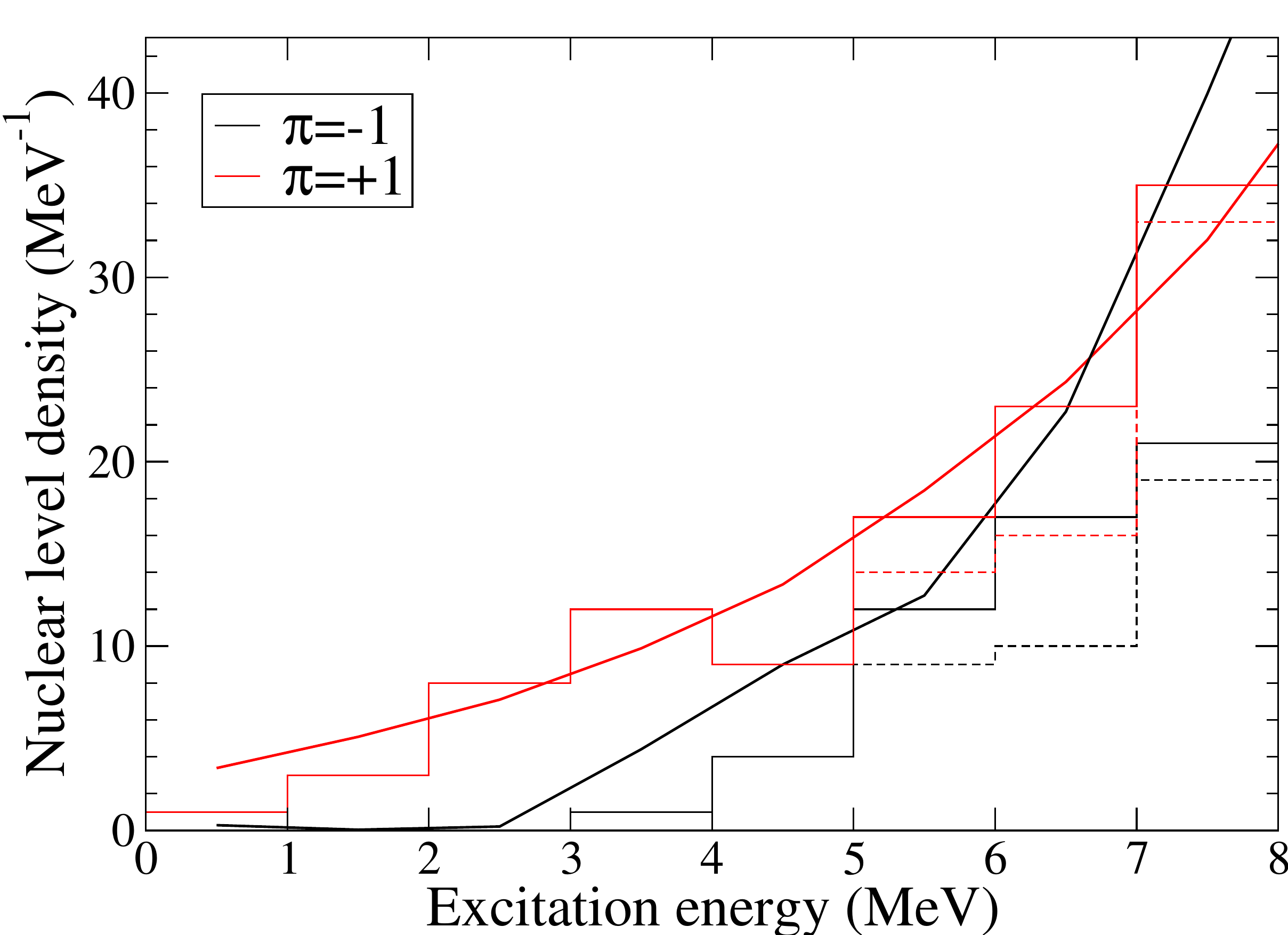}
\end{center}
\caption{Nuclear-level densities for positive parity (red curve) and negative
parity (black curve) of $^{26}$Al compared with experimental data; the solid
and dotted staircases represent upper and lower limits, respectively.
Positive parity NLD is larger than negative parity NLD.} 
\label{fig:levelden}
\end{figure}

Figure~\ref{fig:levelden} shows positive- and negative-parity NLD  for $^{26}$Al
calculated with {\JMOMENTS} compared
with the available experimental data obtained by level counting. Some known
levels have no clear 
assignment of the parity, which leads to upper and lower limits. The calculated
positive-parity NLD is not new, an $sd$-shell calculation being available for
some time. However, the negative-parity NLD was calculated only recently by our
approach \cite{sen'kov11}.

\section{Uncertainty Quantification \pagebudget{(Stefan, Markus, Nicolas, Witek)
(3/4 page)}}
\label{sec:uq}

\begin{figure}[t!]
\includegraphics[width=0.95\linewidth]{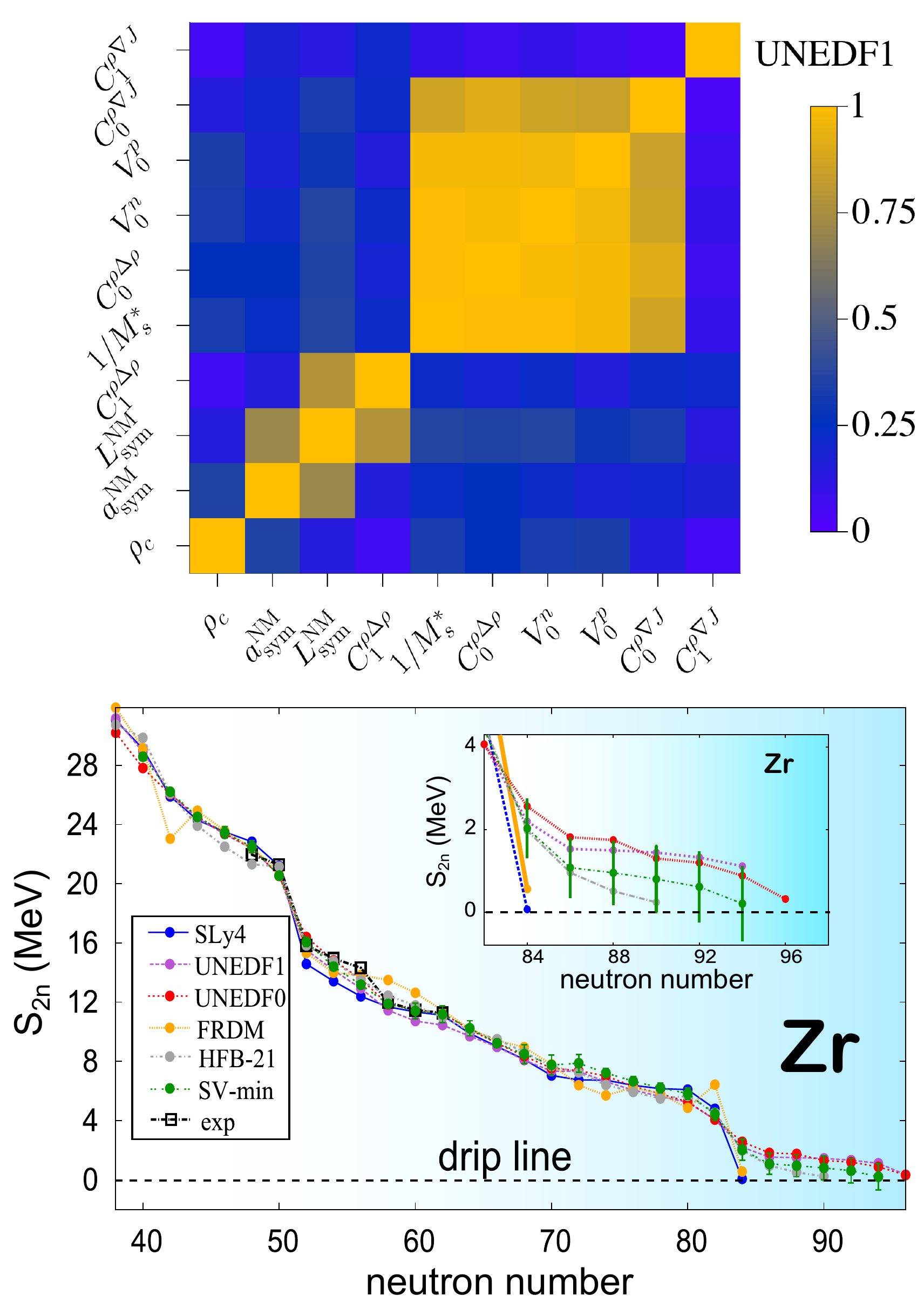}
\caption{
Top: {\UNEDFONE} correlation matrix. Presented are the absolute values
of the correlation coefficients between  the parameters characterizing the
energy density (\ref{eq:EDF}). 
Bottom: Theoretical extrapolations toward drip lines for the
two-neutron separation energies $S_{2n}$ for the isotopic chain of
even-even Zr isotopes using different EDFs ({\sc sl}y4, {\sc sv}-min,
{\UNEDFZERO}, {\UNEDFONE}) \cite{erler2012} and {\sc frdm} \cite{[Moe95]} and
{\sc hfb}-21 \cite{[Gor10]} mass
models.  Detailed predictions around $S_{2n} = 0$
are illustrated in the inset. The
bars on the {\sc sv}-min results indicate statistical errors due to uncertainty
in the
coupling constants of the functional.
} 
\label{fig:uq}
\end{figure}

Uncertainty quantification is a key element for assessing the predictive power
of a model.
When working with effective theories with degrees 
of freedom relevant to the problem, the parameters of the theoretical model 
often need to be adjusted to the empirical input. To quantify the model
uncertainties, 
sensitivity analysis yields the standard deviations and correlations of the
model 
parameters, usually encoded as a covariance matrix
\cite{[Ber05],[Toi08],[Klu09],Rei10}. 

The calculation of the covariance matrix requires computing derivatives of
the 
observables with respect to the model parameters. When a closed-form expression
for the 
derivatives is not available, we estimate the derivatives numerically using
finite differences. To
account for the numerical uncertainty associated with the underlying DFT-based
calculations, we compute the ``noise level'' of each observable following the
approach in \cite{more2010ecn}. The difference parameters used for estimating
the Jacobian matrix associated with (\ref{eq:chi2}) are then obtained using
these noise levels \cite{more2011edn}.

Uncertainty quantification was one of the key topics of the EDF optimization
work performed in the UNEDF collaboration \cite{[Kor10],[Kor12]}. 
The upper panel of  
\Fig{fig:uq} shows the {\UNEDFONE} correlation matrix, obtained from the 
sensitivity analysis. As can be seen, some of the surface parameters of the 
{\UNEDFONE} EDF are strongly correlated. In \cite{erler2012} we used this
information to assess 
the robustness of the current EDFs in the predictions of the nuclear landscape 
limits. This is illustrated in the lower panel of \Fig{fig:uq}, which shows
calculated and experimental two-neutron separation energies for the isotopic
chain of
even-even zirconium isotopes. 
The differences between model predictions are
small in the region where data exist and grow
steadily when extrapolating toward the two-neutron drip line ($S_{2n} = 0$). 
Nevertheless, the consistency between the models was found to 
be surprisingly good. This study required massive parallel calculations of the 
nuclear mass tables \cite{[Erl11]}.

\section{High-Performance Computing Resources}
\label{sec:performance}
UNEDF science has benefited from access to some of the largest computers
in the world, provided primarily by DOE's Innovative and Novel
Computational Impact on Theory and Experiment 
(INCITE) program~\cite{incite-home}. In particular, the largest
computations of UNEDF were carried out on the ``Jaguar'' machine at Oak
Ridge National Laboratory and the ``Intrepid'' machine at Argonne
National Laboratory.
Jaguar has gone through several processor upgrades during the project, taking it
from 30,976 cores (Cray XT4 in 2008) to 298,592 cores (Cray XK6 in 2012); 
Intrepid is an IBM Blue Gene/P with 163,840 processing cores.

\begin{figure}[h!]
\includegraphics[width=1.0\linewidth]{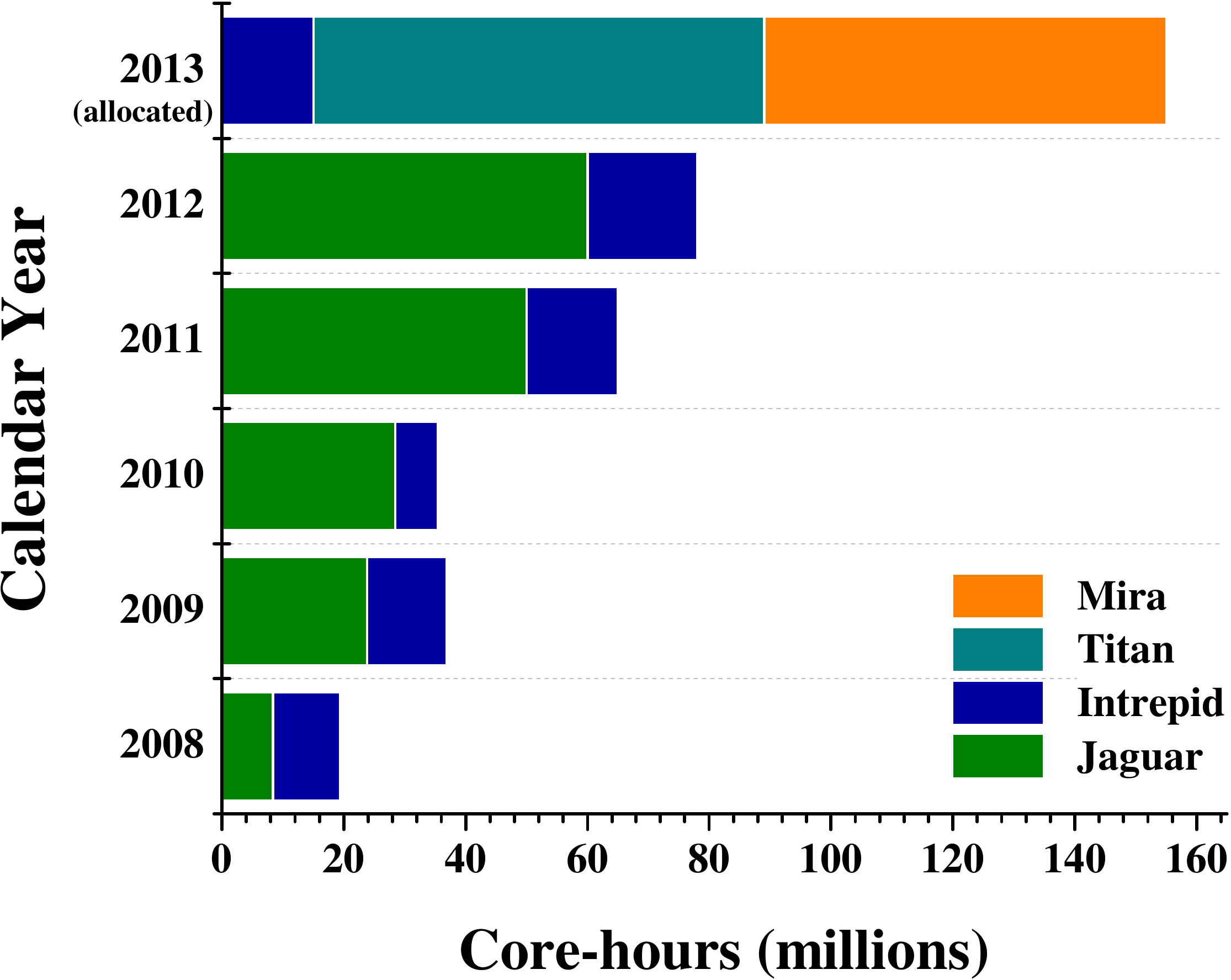}
\caption{UNEDF allocation and utilization (in millions of core-hours) of 
leadership-class computing resources from 2008 to 2013.
}
\label{fig:incite}
\end{figure}

Figure~\ref{fig:incite} shows the UNEDF utilization of these
computing resources over the years 2008-2013 provided through INCITE.
The figure highlights the
increasing demand for computing time in low-energy nuclear physics research. 
The combined 2008 INCITE utilization across Jaguar and Intrepid was nearly 20
million core-hours and by 2012 had increased fourfold.  This growth
illustrates the increasing application of high-performance computing in
nuclear theory enabled by the physics/computer science/applied mathematics 
collaborations fostered by UNEDF.

For the 2013 calendar year, members of the SciDAC-3 NUCLEI project~\cite{NUCLEI}
were granted the sixth largest
allocation of the 61 INCITE projects awarded, with a total allocation of 155
million core-hours across three leadership-class computing resources, Titan,
Mira, and Intrepid. 
Titan is a Cray XK7, a hybrid CPU-GPU system with 299,008 CPU cores and 261,632
GPU streaming multiprocessors, and Mira is an IBM Blue Gene/Q with 786,432
processing cores.  The substantial changes to computing systems at both Argonne
and Oak Ridge, indicative of future trends in high-performance computing, create
new computational
challenges but also new possibilities to achieve larger and more accurate
calculations.  Through the close collaborations enabled through UNEDF, and now
NUCLEI, members are working to continuously scale codes to increase physics
capabilities and improve performance for efficient utilization of these
leadership-class resources.

\section{Conclusions  \pagebudget{(Joe, Rusty, Witek) (1/2 page)}}
\label{sec:conclusion}

The examples presented here illustrate the multifaceted outcomes of the UNEDF
project, both in terms of landmark calculations of nuclear structure and
reactions and in terms of how nuclear theory is done.
The project was very productive,
as can be assessed by going to the project's website, \url{http://unedf.org},
which documents the concrete deliverables of UNEDF: publications, highlights,
reports,
conference presentations, and computer codes.
 UNEDF also placed great importance on recruiting the next
generation of scientists.  Annually it provided training to 30 young
researchers.  The UNEDF experience has been a springboard for advancement, with
many UNEDF postdocs obtaining permanent positions at U.S.\
universities, national laboratories, and overseas institutions.

\begin{figure}[tb]
\includegraphics[width=0.99\linewidth]{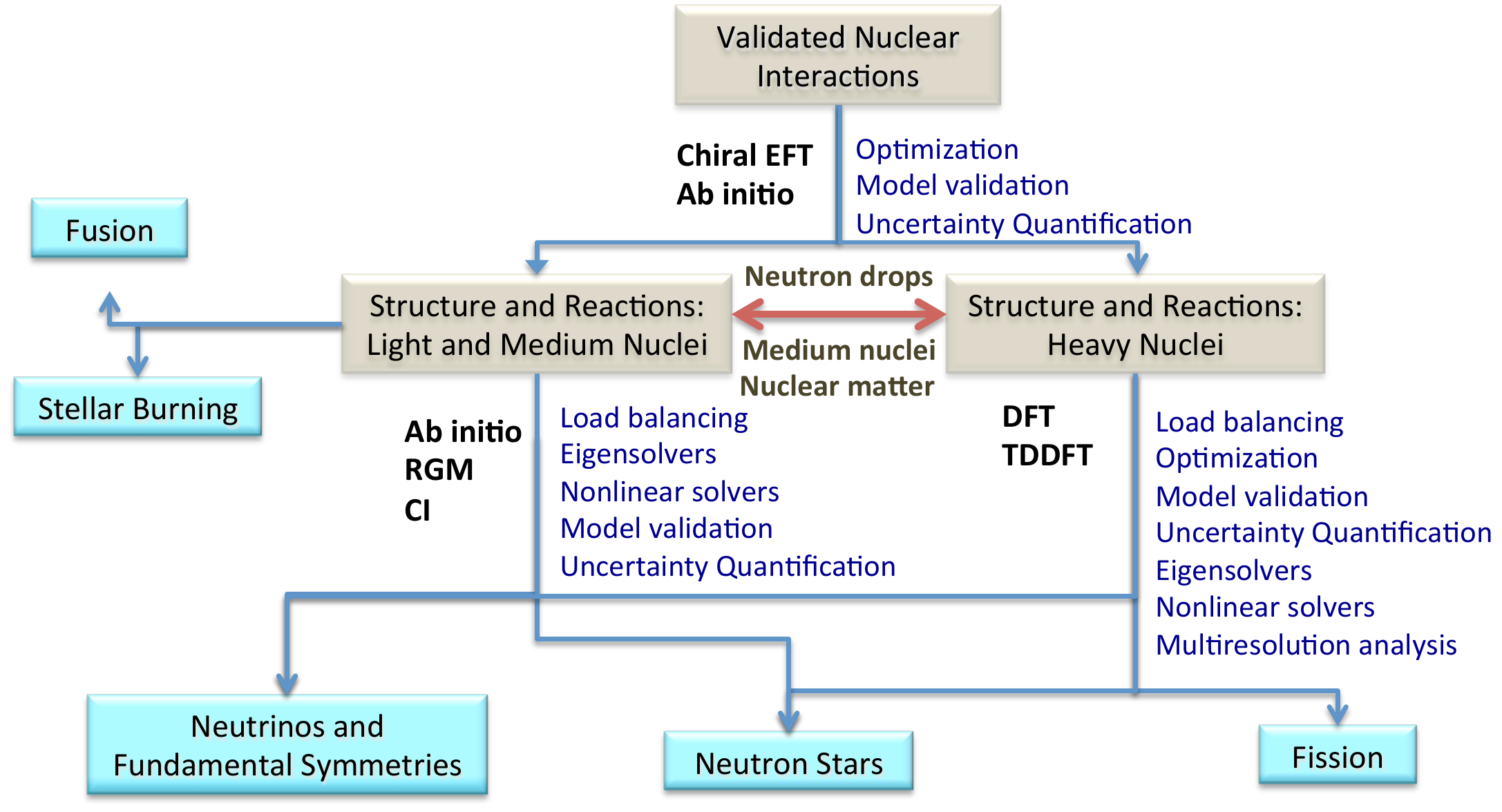}
\caption{Physics and computing in NUCLEI. The major areas of research are
marked, together with connections between them and theoretical and computational
tools. For more details, see \cite{NUCLEI}.} 
\label{fig:nuclei}
\end{figure}

By fostering broad new collaborative efforts between physicists, mathematicians,
and computer scientists, the SciDAC-2 UNEDF project showed how to tackle 
scientific, algorithmic, and computational challenges in the era of
extreme-scale scientific computing.
This effort continues with the SciDAC-3 NUCLEI project \cite{NUCLEI}, which
builds on the
successful  strategies of UNEDF. Figure~\ref{fig:nuclei} shows the key elements
of NUCLEI.

\section*{Acknowledgments}
\label{sec:acknowledgments}

\begin{sloppypar} % This is to fit the long grant numbers
Support for the UNEDF and NUCLEI collaborations was provided through
the SciDAC program funded by the U.S.\ Dept.\ of Energy (DOE), Office of
Science, Advanced Scientific Computing Research and Nuclear Physics programs. 
This work was also supported by DOE Contract Nos.\ 
DE-FG02\--96ER40963 (Univ. Tenn.),
DE-AC52\--07NA27344 (LLNL),
DE-AC02\--05CH11231 (LBNL),
DE-AC05\--00OR22725 (ORNL), 
DE-AC02\--06CH11357 and DE-FC02\--07ER41457 (ANL),
DE-FC02\--09ER41584 (Central Michigan Univ.),
DE-FC02\--09ER41582 (Iowa State Univ.),
DE-FG02\--87ER40371 (Iowa State Univ.),
and
DE-FC02\--09ER41586 (Ohio State Univ.). 
This research used the computational resources of the Oak Ridge Leadership
Computing Facility (OLCF) at ORNL and Argonne Leadership Computing Facility
(ALCF) at ANL provided through the INCITE program. Computational resources were
also provided by the National Institute for Computational Sciences (NICS) at
ORNL, the Laboratory Computing Resource Center (LCRC) at ANL, and the National
Energy Research Scientific Computing Center (NERSC) at LBNL.
\end{sloppypar}

\bibliographystyle{elsarticle-num}
\bibliography{master}

\end{document}

%% file: preamble.tex
\usepackage{graphicx}
\usepackage{amssymb}
\usepackage{amsthm}

\usepackage{url}
\usepackage{color}
\usepackage{aas_macros}	% Defines bib macros
\newcommand{\pagebudget}[1]{ } % In final version
\usepackage{comment} % For comment environment

\newcommand{\POUNDERS}{\textsc{pound}er\textsc{s}}
\newcommand{\ADLB}{\textsc{adlb}}

\newcommand{\AGFMC}{\textsc{agfmc}}

\newcommand{\ignore}[1]{}  % Use to comment out large blocks of text
\newcommand{\Sec}[1]{{Sec.~\ref{#1}}}
\newcommand{\Fig}[1]{Fig.~\ref{#1}}
\newcommand{\R}{\mathbb R}    % Real numbers
\newcommand{\xb}{x}  % bold for vectors
\newcommand{\gras}[1]{\boldsymbol{#1}} % bold font for symbols
\newcommand{\UNEDFZERO}{\textsc{unedf0}}

\newcommand{\UNEDFONE}{\textsc{unedf1}}
\newcommand{\UNEDFTWO}{\textsc{unedf2}}
\newcommand{\HFBTHO}{\textsc{hfbtho}}
\newcommand{\HFODD}{\textsc{hfodd}}
\newcommand{\MFDn}{\textsc{mfd}n}
\newcommand{\JMOMENTS}{\textsc{jmoments}}
\newcommand{\MFDnJ}{\textsc{mfdnj}}
\newcommand{\BIGSTICK}{\textsc{bigstick}}
\newcommand{\NuShellX}{\textsc{nushellx}}
\newcommand{\HFBRAD}{\textsc{hfbrad}}
\newcommand{\HFBMARIO}{\textsc{hfbmario}}
\newcommand{\HFBAX}{\textsc{hfb-ax}}
\newcommand{\HFBMAD}{\textsc{madness-hfb}}

\newcommand{\LAPACK}{\textsc{lapack}}
\newcommand{\SCALAPACK}{\textsc{scalapack}}
\newcommand{\BLAS}{\textsc{blas}}

\usepackage{bm}% bold math

%% file: main_tr.bbl
\begin{thebibliography}{100}
\expandafter\ifx\csname url\endcsname\relax
  \def\url#1{\texttt{#1}}\fi
\expandafter\ifx\csname urlprefix\endcsname\relax\def\urlprefix{URL }\fi
\expandafter\ifx\csname href\endcsname\relax
  \def\href#1#2{#2} \def\path#1{#1}\fi

\bibitem{Bertsch:2007}
G.~F. {Bertsch}, D.~J. {Dean}, W.~{Nazarewicz}, \srev\ 6 (2007) 42.

\bibitem{Furn:2011}
R.~Furnstahl, \npn\ 21 (2011) 24.

\bibitem{GFMC1987}
J.~Carlson, \prc\ 36 (1987) 2026.
\newblock

\bibitem{Pieper:2001mp}
S.~C. Pieper, R.~B. Wiringa, \arnps\ 51 (2001) 53.
\newblock

\bibitem{GFMCscidac:2010aa}
E.~Lusk, S.~C. Pieper, R.~Butler, \srev\ 17 (2010) 30.

\bibitem{Nol:2007}
K.~M. Nollett, S.~C. Pieper, R.~B. Wiringa, J.~Carlson, G.~M. Hale, \prl\ 99
  (2007) 022502.
\newblock

\bibitem{Pud:1997}
B.~S. Pudliner, V.~R. Pandharipande, J.~Carlson, S.~C. Pieper, R.~B. Wiringa,
  \prc\ 56 (1997) 1720.
\newblock

\bibitem{Wir:2000}
R.~B. Wiringa, S.~C. Pieper, J.~Carlson, V.~R. Pandharipande, \prc\ 62 (2000)
  014001.
\newblock

\bibitem{Pie:2004}
S.~C. Pieper, R.~B. Wiringa, J.~Carlson, \prc\ 70 (2004) 054325.
\newblock

\bibitem{Epe:2011}
E.~{Epelbaum}, H.~{Krebs}, D.~{Lee}, U.-G. {Mei{\ss}ner}, \prl\ 106 (2011)
  192501.
\newblock

\bibitem{Epelbaum:2012qn}
E.~Epelbaum, H.~Krebs, T.~A. Lahde, D.~Lee, U.-G. Mei{\ss}ner, \prl\ 109 (2012)
  252501.
\newblock

\bibitem{Wir:2008}
R.~B. {Wiringa}, R.~{Schiavilla}, S.~C. {Pieper}, J.~{Carlson}, \prc\ 78 (2008)
  021001.

\bibitem{Per:2007}
M.~{Pervin}, S.~C. {Pieper}, R.~B. {Wiringa}, \prc\ 76 (2007) 064319.
\newblock

\bibitem{aroua}
S.~Aroua, P.~Navr\'atil, L.~Zamick, M.~S. Fayache, B.~R. Barrett, J.~P. Vary,
  K.~Heyde, \nphysa\ 720 (2011) 71.
\newblock

\bibitem{Maris:2011as}
P.~Maris, J.~P. Vary, P.~Navratil, W.~E. Ormand, H.~Nam, D.~J. Dean, \prl\ 106
  (2011) 202502.
\newblock

\bibitem{SternbergSC08}
P.~Sternberg, E.~G. Ng, C.~Yang, P.~Maris, J.~P. Vary, M.~Sosonkina, H.~V. Le,
  Accelerating configuration interaction calculations for nuclear structure,
  in: Proc. 2008 ACM/IEEE Conf. Supercomputing, 2008.

\bibitem{VarySciDAC09}
J.~P. Vary, P.~Maris, E.~Ng, C.~Yang, M.~Sosonkina, \jpcs\ 180 (2009) 012083.
\newblock

\bibitem{MarisICCS10}
P.~Maris, M.~Sosonkina, J.~P. Vary, E.~Ng, C.~Yang, \pcs\ 1 (2010) 97.
\newblock

\bibitem{Aktulga_new}
H.~Aktulga, C.~Yang, E.~Ng, P.~Maris, J.~Vary, LCNS 7484 (2012) 830.
\newblock

\bibitem{holtjw2008}
J.~W. Holt, G.~E. Brown, T.~T.~S. Kuo, J.~D. Holt, R.~Machleidt, \prl\ 100
  (2008) 062501.
\newblock

\bibitem{NCSM_review}
B.~R. Barrett, P.~{Navr\'atil}, J.~P. Vary, \ppnp\ 69 (2013) 131.
\newblock

\bibitem{Maris:2012du}
P.~Maris, H.~M. Aktulga, M.~A. Caprio, U.~Catalyurek, E.~G. Ng, et~al., \jpcs\
  403 (2012) 012019.
\newblock

\bibitem{Maris:2009bx}
P.~Maris, A.~Shirokov, J.~Vary, \prc\ C81 (2010) 021301.
\newblock

\bibitem{Maris:2013rgq}
P.~Maris, J.~P. Vary, S.~Gandolfi, J.~Carlson, S.~C. Pieper, arXiv:1302.2089.

\bibitem{Qua:2008}
S.~Quaglioni, P.~{Navr{\'a}til}, \prl\ 101 (2008) 092501.
\newblock

\bibitem{Qua:2009}
S.~Quaglioni, P.~Navr\'atil, \prc\ 79 (2009) 044606.
\newblock

\bibitem{PRL110}
S.~Baroni, P.~{Navr\'atil}, S.~Quaglioni, \prl\ 110 (2013) 022505.
\newblock

\bibitem{NCSMRG2012:navratil}
P.~Navr\'atil, S.~Quaglioni, \prl\ 108 (2012) 042503.
\newblock

\bibitem{Navratil:2011ab}
P.~Navr\'atil, R.~Roth, S.~Quaglioni, \plettb\ 704 (2011) 379.
\newblock

\bibitem{coester1958}
F.~Coester, \nphys\ 7 (1958) 421.
\newblock

\bibitem{coester1960}
F.~Coester, H.~K{\"u}mmel, \nphys\ 17 (1960) 477.
\newblock

\bibitem{kuemmel1978}
H.~K{\"u}mmel, K.~H. L{\"u}hrmann, J.~G. Zabolitzky, \physrep\ 36 (1978) 1.
\newblock

\bibitem{bartlett2007}
R.~J. Bartlett, M.~Musia\l{}, \revmp\ 79 (2007) 291.
\newblock

\bibitem{dean2004}
D.~J. Dean, M.~Hjorth-Jensen, \prc\ 69 (2004) 054320.
\newblock

\bibitem{hagen2010b}
G.~Hagen, T.~Papenbrock, D.~J. Dean, M.~Hjorth-Jensen, \prc\ 82 (2010) 034330.
\newblock

\bibitem{entem2003}
D.~R. Entem, R.~Machleidt, \prc\ 68 (2003) 041001.
\newblock

\bibitem{hagen2008}
G.~Hagen, T.~Papenbrock, D.~J. Dean, M.~Hjorth-Jensen, \prl\ 101 (2008) 092502.
\newblock

\bibitem{hagen2007d}
G.~Hagen, D.~J. Dean, M.~Hjorth-Jensen, T.~Papenbrock, \plettb\ 656 (2007) 169.
\newblock

\bibitem{hagen2010a}
G.~Hagen, T.~Papenbrock, M.~Hjorth-Jensen, \prl\ 104 (2010) 182501.
\newblock

\bibitem{hagen2009a}
G.~Hagen, T.~Papenbrock, D.~J. Dean, \prl\ 103 (2009) 062503.
\newblock

\bibitem{jansen2011}
G.~R. Jansen, M.~Hjorth-Jensen, G.~Hagen, T.~Papenbrock, \prc\ 83 (2011)
  054306.
\newblock

\bibitem{holtjw2009}
J.~W. Holt, N.~Kaiser, W.~Weise, \prc\ 79 (2009) 054331.
\newblock

\bibitem{hebeler2010b}
K.~Hebeler, A.~Schwenk, \prc\ 82 (2010) 014314.
\newblock

\bibitem{hagen2012a}
G.~Hagen, M.~Hjorth-Jensen, G.~R. Jansen, R.~Machleidt, T.~Papenbrock, \prl\
  108 (2012) 242501.
\newblock

\bibitem{furnstahl2012}
R.~J. Furnstahl, G.~Hagen, T.~Papenbrock, \prc\ 86 (2012) 031301.
\newblock

\bibitem{hagen2009b}
G.~Hagen, T.~Papenbrock, D.~J. Dean, M.~Hjorth-Jensen, B.~V. Asokan, \prc\ 80
  (2009) 021306.
\newblock

\bibitem{hagen2012c}
G.~{Hagen}, H.~A. {Nam}, \ptps\ 196 (2012) 102.

\bibitem{hagen2007b}
G.~Hagen, D.~J. Dean, M.~Hjorth-Jensen, T.~Papenbrock, A.~Schwenk, \prc\ 76
  (2007) 044305.
\newblock

\bibitem{hagen2012b}
G.~Hagen, M.~Hjorth-Jensen, G.~R. Jansen, R.~Machleidt, T.~Papenbrock, \prl\
  109 (2012) 032502.
\newblock

\bibitem{prisciandaro2001}
J.~I. Prisciandaro, P.~F. Mantica, B.~A. Brown, D.~W. Anthony, M.~W. Cooper,
  A.~Garcia, D.~E. Groh, A.~Komives, W.~Kumarasiri, P.~A. Lofy, A.~M.
  Oros-Peusquens, S.~L. Tabor, M.~Wiedeking, \plettb\ 510 (2001) 17.
\newblock

\bibitem{janssens2002}
R.~V.~F. Janssens, B.~Fornal, P.~F. Mantica, B.~A. Brown, R.~Broda,
  P.~Bhattacharyya, M.~P. Carpenter, M.~Cinausero, P.~J. Daly, A.~D. Davies,
  T.~Glasmacher, Z.~W. Grabowski, D.~E. Groh, M.~Honma, F.~G. Kondev,
  W.~Kr\'olas, T.~Lauritsen, S.~N. Liddick, S.~Lunardi, N.~Marginean,
  T.~Mizusaki, D.~J. Morrissey, A.~C. Morton, W.~F. Mueller, T.~Otsuka,
  T.~Pawlat, D.~Seweryniak, H.~Schatz, A.~Stolz, S.~L. Tabor, C.~A. Ur,
  G.~Viesti, I.~Wiedenhaver, J.~Wrzesinski, \plettb\ 546 (2002) 55.
\newblock

\bibitem{honma2002}
M.~Honma, T.~Otsuka, B.~A. Brown, T.~Mizusaki, \prc\ 65 (2002) 061301.
\newblock

\bibitem{liddick2004}
S.~N. Liddick, P.~F. Mantica, R.~V.~F. Janssens, R.~Broda, B.~A. Brown, M.~P.
  Carpenter, B.~Fornal, M.~Honma, T.~Mizusaki, A.~C. Morton, W.~F. Mueller,
  T.~Otsuka, J.~Pavan, A.~Stolz, S.~L. Tabor, B.~E. Tomlin, M.~Wiedeking, \prl\
  92 (2004) 072502.
\newblock

\bibitem{dinca2005}
D.-C. Dinca, R.~V.~F. Janssens, A.~Gade, D.~Bazin, R.~Broda, B.~A. Brown, C.~M.
  Campbell, M.~P. Carpenter, P.~Chowdhury, J.~M. Cook, A.~N. Deacon, B.~Fornal,
  S.~J. Freeman, T.~Glasmacher, M.~Honma, F.~G. Kondev, J.-L. Lecouey, S.~N.
  Liddick, P.~F. Mantica, W.~F. Mueller, H.~Olliver, T.~Otsuka, J.~R. Terry,
  B.~A. Tomlin, K.~Yoneda, \prc\ 71 (2005) 041302.
\newblock

\bibitem{vankolck1994}
U.~van Kolck, \prc\ 49 (1994) 2932.
\newblock

\bibitem{epelbaum2009}
E.~Epelbaum, H.-W. Hammer, U.-G. Mei\ss{}ner, \revmp\ 81 (2009) 1773.
\newblock

\bibitem{machleidt2011}
R.~Machleidt, D.~Entem, \physrep\ 503 (2011) 1.
\newblock

\bibitem{holt2012}
J.~D. Holt, T.~Otsuka, A.~Schwenk, T.~Suzuki, \jpg\ 39 (2012) 085111.

\bibitem{steppenbeck2012}
D.~Steppenbeck, private communication (2012).

\bibitem{BIGSTICK}
C.~W. Johnson, W.~E. Ormand, P.~G. Krastev, arXiv:1303.0905.

\bibitem{Caurier1}
E.~Caurier, F.~Nowacki, \actappb\ 30 (1999) 705.

\bibitem{Caurier2}
E.~Caurier, G.~Mart\'inez-Pinedo, F.~Nowacki, A.~Poves, J.~Retamosa, A.~P.
  Zuker, \prc\ 59 (1999) 2033.
\newblock

\bibitem{AktulgaHPCS11}
H.~Aktulga, C.~Yang, E.~Ng, P.~Maris, J.~Vary, Large-scale parallel null space
  calculation for nuclear configuration interaction problem, in: Proc. Intern.
  Conf. on High Performance Computing and Simulation, 2011.
\newblock

\bibitem{AktulgaHPSS11}
H.~Aktulga, C.~Yang, U.~Catalyurek, P.~Maris, J.~Vary, E.~Ng, On reducing {I/O}
  overheads in large-scale invariant subspace projections, in: Proc. 2011
  Workshop on Algorithms and Programming Tools for Next-Generation
  High-Performance Scientific Software, 2011, p. 305.

\bibitem{nsx-ws}
{NuShellX}, see \url{http://www.garsington.eclipse.co.uk}.

\bibitem{Ben03}
M.~Bender, P.-H. Heenen, P.-G. Reinhard, \revmp\ 75 (2003) 121.
\newblock

\bibitem{[Eng75]}
Y.~M. Engel, D.~M. Brink, K.~Goeke, S.~Krieger, D.~Vautherin, \nphysa\ 249
  (1975) 215.

\bibitem{Roh10}
S.~G. Rohozi\'{n}ski, J.~Dobaczewski, W.~Nazarewicz, \prc\ 81 (2010) 014313.
\newblock

\bibitem{Dob84}
J.~Dobaczewski, H.~Flocard, J.~Treiner, \nphysa\ 422 (1984) 103.

\bibitem{Pei11}
J.~C. Pei, A.~T. Kruppa, W.~Nazarewicz, \prc\ 84 (2011) 024311.
\newblock

\bibitem{[Kor10]}
M.~Kortelainen, T.~Lesinski, J.~Mor\'{e}, W.~Nazarewicz, J.~Sarich, N.~Schunck,
  M.~V. Stoitsov, S.~Wild, \prc\ 82 (2010) 024313.
\newblock

\bibitem{[Kor12]}
M.~Kortelainen, J.~McDonnell, W.~Nazarewicz, P.-G. Reinhard, J.~Sarich,
  N.~Schunck, M.~Stoitsov, S.~Wild, \prc\ 85 (2012) 024304.
\newblock

\bibitem{Stoitsov03}
M.~V. Stoitsov, J.~Dobaczewski, W.~Nazarewicz, S.~Pittel, D.~J. Dean, \prc\ 68
  (2003) 054312.

\bibitem{[Sto09a]}
M.~Stoitsov, W.~Nazarewicz, N.~Schunck, \ijmpe\ 18 (2009) 816.
\newblock

\bibitem{[Erl11]}
J.~Erler, N.~Birge, M.~Kortelainen, W.~Nazarewicz, E.~Olsen, A.~M. Perhac,
  M.~Stoitsov, \jpcs\ 402 (2012) 012030.
\newblock

\bibitem{erler2012}
J.~Erler, N.~Birge, M.~Kortelainen, W.~Nazarewicz, E.~Olsen, A.~M. Perhac,
  M.~Stoitsov, \nat\ 486 (2012) 509.
\newblock

\bibitem{Sto05}
M.~Stoitsov, J.~Dobaczewski, W.~Nazarewicz, P.~Ring, \cpc\ 167 (2005) 43.
\newblock

\bibitem{Dob97}
J.~Dobaczewski, J.~Dudek, \cpc\ 102 (1997) 183.

\bibitem{Sch13}
M.~Stoitsov, N.~Schunck, M.~Kortelainen, N.~Michel, H.~Nam, E.~Olsen,
  J.~Sarich, S.~Wild, \cpc\ 184 (2013) 1592.
\newblock

\bibitem{Sch11}
N.~Schunck, J.~Dobaczewski, J.~McDonnell, W.~Satu{\l}a, J.~Sheikh,
  A.~Staszczak, M.~Stoitsov, P.~Toivanen, \cpc\ 183 (2012) 166.
\newblock

\bibitem{Sta09}
A.~Staszczak, A.~Baran, J.~Dobaczewski, W.~Nazarewicz, \prc\ 80 (2009) 014309.
\newblock

\bibitem{[Yue12]}
Y.~Shi, J.~Dobaczewski, S.~Frauendorf, W.~Nazarewicz, J.~Pei, F.~Xu,
  N.~Nikolov, \prl\ 108 (2012) 092501.

\bibitem{Sch10}
N.~Schunck, J.~Dobaczewski, J.~McDonnell, J.~Mor\'e, W.~Nazarewicz, J.~Sarich,
  M.~V. Stoitsov, \prc\ 81 (2010) 024316.
\newblock

\bibitem{Sta10}
A.~Staszczak, M.~Stoitsov, A.~Baran, W.~Nazarewicz, \epja\ 46 (2010) 85.
\newblock

\bibitem{[Bar08]}
A.~Baran, A.~Bulgac, M.~Forbes, G.~Hagen, W.~Nazarewicz, N.~Schunck, M.~V.
  Stoitsov, \prc\ 78 (2008) 014318.
\newblock

\bibitem{Pei08}
J.~C. Pei, M.~V. Stoitsov, G.~I. Fann, W.~Nazarewicz, N.~Schunck, F.~R. Xu,
  \prc\ 78 (2008) 064306.
\newblock

\bibitem{[Pei09]}
J.~Pei, W.~Nazarewicz, J.~Sheikh, A.~Kerman, \prl\ 102 (2009) 192501.
\newblock

\bibitem{Pei10}
J.~C. Pei, J.~Dukelsky, W.~Nazarewicz, \pra\ 82 (2010) 021603.

\bibitem{Par13}
R.~Parrish, E.~Hoenstein, N.~Schunck, C.~Sherril, T.~Martinez, arXiv:1301.5064.

\bibitem{ABGV02}
B.~Alpert, G.~Beylkin, D.~Gines, L.~Vozovoi, \jcomp\ 182 (2002) 149.

\bibitem{A-B-C-R:1993}
B.~Alpert, G.~Beylkin, R.~Coifman, V.~Rokhlin, \sisc\ 14 (1993) 159.

\bibitem{BR-FA-YA:1997}
M.~Brewster, G.~Fann, Z.~Yang, \jmc\ 22 (1997) 117.

\bibitem{FBHJ04}
G.~Fann, G.~Beylkin, R.~Harrison, K.~Jordan, \ibmjrd\ 48 (2004) 161.

\bibitem{YFGHB04-2}
T.~Yanai, G.~I. Fann, Z.~Gan, R.~J. Harrison, G.~Beylkin, \jcp\ 121 (2004)
  6680.

\bibitem{BCFH07}
G.~Beylkin, R.~Cramer, G.~Fann, R.~J. Harrison, \jacha\ 23 (2007) 235.

\bibitem{PhysRevA.85.033403}
N.~Vence, R.~Harrison, P.~Krsti\ifmmode~\acute{c}\else \'{c}\fi{}, \pra\ 85
  (2012) 033403.
\newblock

\bibitem{JHF2011}
J.~Jia, R.~J. Harrison, G.~Fann, submitted.

\bibitem{Ben05}
K.~Bennaceur, J.~Dobaczewski, \cpc\ 168 (2005) 96.
\newblock

\bibitem{FPHJHONSS09}
G.~I. Fann, J.~Pei, R.~J. Harrison, J.~Jia, J.~Hill, M.~Ou, W.~Nazarewicz,
  W.~A. Shelton, N.~Schunck, \jpcs\ 180.
\newblock

\bibitem{[Nik11]}
N.~Nikolov, N.~Schunck, W.~Nazarewicz, M.~Bender, J.~Pei, \prc\ 83 (2011)
  034305.
\newblock

\bibitem{tao-man}
T.~Munson, J.~Sarich, S.~M. Wild, S.~Benson, L.~{Curfman McInnes}, {TAO} 2.0
  users manual, Technical Memorandum ANL/MCS-TM-322, Argonne National
  Laboratory, Argonne, Illinois, see \url{http://www.mcs.anl.gov/tao} (2012).

\bibitem{Gandolfiprc:2012}
S.~{Gandolfi}, J.~{Carlson}, S.~{Reddy}, \prc\ 85 (2012) 032801.
\newblock

\bibitem{Tsangprc:2012}
M.~B. Tsang, J.~R. Stone, F.~Camera, P.~Danielewicz, S.~Gandolfi, K.~Hebeler,
  C.~J. Horowitz, J.~Lee, W.~G. Lynch, Z.~Kohley, R.~Lemmon, P.~M\"oller,
  T.~Murakami, S.~Riordan, X.~Roca-Maza, F.~Sammarruca, A.~W. Steiner,
  I.~Vida\~na, S.~J. Yennello, \prc\ 86 (2012) 015803.
\newblock

\bibitem{Gandolfimnras:2010}
S.~Gandolfi, A.~Y. Illarionov, S.~Fantoni, J.~Miller, F.~Pederiva, K.~Schmidt,
  \mnras\ 404 (2010) L35.
\newblock

\bibitem{Gandolfi:2009}
S.~{Gandolfi}, A.~Y. {Illarionov}, K.~E. {Schmidt}, F.~{Pederiva},
  S.~{Fantoni}, \prc\ 79 (2009) 054005.
\newblock

\bibitem{Chang:2004}
S.-Y. {Chang}, J.~{Morales}, V.~R. {Pandharipande}, D.~G. {Ravenhall},
  J.~{Carlson}, S.~C. {Pieper}, R.~B. {Wiringa}, K.~E. {Schmidt}, \nphysa\ 746
  (2004) 215.
\newblock

\bibitem{Gandolfi:2006}
S.~Gandolfi, F.~Pederiva, S.~Fantoni, K.~E. Schmidt, \prc\ 73 (2006) 044304.
\newblock

\bibitem{Gandolfiepja:2008}
S.~{Gandolfi}, F.~{Pederiva}, S.~{A Beccara}, \epja\ 35 (2008) 207.
\newblock

\bibitem{Gan11b}
S.~Gandolfi, J.~Carlson, S.~C. Pieper, \prl\ 106 (2011) 012501.
\newblock

\bibitem{Steiner:2012}
A.~W. Steiner, S.~Gandolfi, \prl\ 108 (2012) 081102.
\newblock

\bibitem{Sto10}
M.~Stoitsov, M.~Kortelainen, S.~K. Bogner, T.~Duguet, R.~J. Furnstahl,
  B.~Gebremariam, N.~Schunck, \prc\ 82 (2010) 054307.
\newblock

\bibitem{Bog11}
S.~K. Bogner, R.~J. Furnstahl, H.~Hergert, M.~Kortelainen, P.~Maris,
  M.~Stoitsov, J.~P. Vary, \prc\ 84 (2011) 044306.
\newblock

\bibitem{Drut:2010}
J.~E. {Drut}, R.~J. {Furnstahl}, L.~{Platter}, \ppnp\ 64 (2010) 120.
\newblock

\bibitem{Jurgenson:2009qs}
E.~D. Jurgenson, P.~Navratil, R.~J. Furnstahl, \prl\ 103 (2009) 082501.
\newblock

\bibitem{Bogner:2009bt}
S.~K. Bogner, R.~J. Furnstahl, A.~Schwenk, \ppnp\ 65 (2010) 94.
\newblock

\bibitem{jurgenson2011}
E.~D. Jurgenson, P.~Navr\'atil, R.~J. Furnstahl, \prc\ 83 (2011) 034301.
\newblock

\bibitem{roth2011a}
R.~Roth, J.~Langhammer, A.~Calci, S.~Binder, P.~Navr\'atil, \prl\ 107 (2011)
  072501.
\newblock

\bibitem{roth2012}
R.~Roth, S.~Binder, K.~Vobig, A.~Calci, J.~Langhammer, P.~Navr\'atil, \prl\ 109
  (2012) 052501.
\newblock

\bibitem{Heb10}
K.~Hebeler, S.~Bogner, R.~Furnstahl, A.~Nogga, A.~Schwenk, \prc\ 83 (2011)
  031301.
\newblock

\bibitem{Negele:1972zp}
J.~W. Negele, D.~Vautherin, \prc\ 5 (1972) 1472.

\bibitem{Bogner:2008kj}
S.~K. Bogner, R.~J. Furnstahl, L.~Platter, \epja\ 39 (2009) 219.
\newblock

\bibitem{Geb10}
B.~Gebremariam, T.~Duguet, S.~K. Bogner, \prc\ 82 (2010) 014305.
\newblock

\bibitem{Carlsson:2010da}
B.~Carlsson, J.~Dobaczewski, \prl\ 105 (2010) 122501.
\newblock

\bibitem{Dobaczewski:2010qp}
J.~Dobaczewski, B.~Carlsson, M.~Kortelainen, \jpg\ 37 (2010) 075106.
\newblock

\bibitem{Ter05}
J.~Terasaki, J.~Engel, M.~Bender, J.~Dobaczewski, W.~Nazarewicz, M.~Stoitsov,
  \prc\ 71 (2005) 034310.
\newblock

\bibitem{Ter10}
J.~Terasaki, J.~Engel, \prc\ 82 (2010) 034326.
\newblock

\bibitem{[Yosh09]}
K.~Yoshida, \prc\ 80 (2009) 044324.

\bibitem{[Ebat10]}
S.~Ebata, T.~Nakatsukasa, T.~Inakura, K.~Yoshida, Y.~Hashimoto, K.~Yabana,
  \prc\ 82 (2010) 034306.

\bibitem{[Pena09]}
D.~P. Arteaga, E.~Khan, P.~Ring, \prc\ 79 (2009) 034311.

\bibitem{[Peru11]}
S.~Peru, G.~Gosselin, M.~Martini, M.~Dupuis, S.~Hilaire, J.-C. Devaux, \prc\ 83
  (2011) 014314.

\bibitem{[Losa10]}
C.~Losa, A.~Pastore, T.~Dossing, E.~Vigezzi, R.~Broglia, \prc\ 81 (2010)
  064307.

\bibitem{Rin80}
P.~Ring, P.~Schuck, The Nuclear Many-body Problem, Springer-Verlag, Berlin,
  1980.

\bibitem{Bla05}
A.~Blazkiewicz, V.~E. Oberacker, A.~S. Umar, M.~Stoitsov, \prc\ 71 (2005)
  054321.
\newblock

\bibitem{Ter11}
J.~Terasaki, J.~Engel, \prc\ 84 (2011) 014332.
\newblock

\bibitem{PhysRevLett.105.202502}
G.~P.~A. Nobre, F.~S. Dietrich, J.~E. Escher, I.~J. Thompson, M.~Dupuis,
  J.~Terasaki, J.~Engel, \prl\ 105 (2010) 202502.
\newblock

\bibitem{PhysRevC.84.064609}
G.~P.~A. Nobre, F.~S. Dietrich, J.~E. Escher, I.~J. Thompson, M.~Dupuis,
  J.~Terasaki, J.~Engel, \prc\ 84 (2011) 064609.
\newblock

\bibitem{JourPhys.312.082033}
G.~P.~A. Nobre, I.~J. Thompson, J.~E. Escher, F.~S. Dietrich, \jpcs\ 312 (2011)
  082033.
\newblock

\bibitem{Koning2003NPA713}
A.~J. Koning, J.~P. Delaroche, \nphysa\ 713 (2003) 231.
\newblock

\bibitem{PhysRevC.4.1114}
J.~J.~H. Menet, E.~E. Gross, J.~J. Malanify, A.~Zucker, \prc\ 4 (1971) 1114.
\newblock

\bibitem{[Avog11]}
P.~Avogadro, T.~Nakatsukasa, \prc\ 84 (2011) 014314.

\bibitem{[Avog13]}
P.~Avogadro, T.~Nakatsukasa, \prc\ 87 (2013) 014331.

\bibitem{[Stoi11]}
M.~Stoitsov, M.~Kortelainen, T.~Nakatsukasa, C.~Losa, W.~Nazarewicz, \prc\ 84
  (2011) 041305(R).

\bibitem{Toivanen10}
J.~Toivanen, B.~G. Carlsson, J.~Dobaczewski, K.~Mizuyama, R.~R.
  Rodr\'\i{}guez-Guzm\'an, P.~Toivanen, P.~Vesel\'y, \prc\ 81 (2010) 034312.

\bibitem{Carlsson12}
B.~G. Carlsson, J.~Toivanen, A.~Pastore, \prc\ 86 (2012) 014307.

\bibitem{BY:2002fk}
A.~Bulgac, Y.~Yu, \prl\ 88 (2002) 042504.
\newblock

\bibitem{Bulgac:2002uq}
A.~Bulgac, \prc\ 65 (2002) 051305(R).
\newblock

\bibitem{Bulgac:2007a}
A.~Bulgac, \pra\ 76 (2007) 040502.
\newblock

\bibitem{BF:2008}
A.~Bulgac, M.~M. Forbes, \prl\ 101 (2008) 215301.
\newblock

\bibitem{Bulgac:2009b}
A.~Bulgac, S.~Yoon, \prl\ 102 (2009) 085302.
\newblock

\bibitem{Bulgac:2011}
A.~Bulgac, M.~M. Forbes, P.~Magierski, The Unitary Fermi Gas: From {Monte
  Carlo} to Density Functionals, Vol. 836, Springer-Verlag, Berlin, 2012,
  Ch.~9, p. 305.

\bibitem{Bulgac:2011b}
A.~Bulgac, Y.-L. Luo, P.~Magierski, K.~J. Roche, Y.~Yu, \sci\ 332 (2011) 1288.
\newblock

\bibitem{Bulgac:2011c}
A.~Bulgac, Y.-L. Luo, K.~J. Roche, \prl\ 108 (2012) 150401.
\newblock

\bibitem{Stetcu:2011}
I.~Stetcu, A.~Bulgac, P.~Magierski, K.~J. Roche, \prc\ 84 (2011) 051309(R).
\newblock

\bibitem{Bulgac:2013b}
A.~Bulgac, arXiv:1301.0357.

\bibitem{Bulgac:2008}
A.~Bulgac, K.~J. Roche, \jpcs\ 125 (2008) 012064.
\newblock

\bibitem{Bulgac:2013}
A.~Bulgac, M.~M. Forbes, arXiv:1301.7354.

\bibitem{Hauser52}
W.~Hauser, H.~Feshbach, \pr\ 87 (1952) 366.
\newblock

\bibitem{scott10}
M.~Scott, M.~Horoi, \epl\ 91 (2010) 52001.
\newblock

\bibitem{horoi07}
M.~Horoi, V.~Zelevinsky, \prl\ 98 (2007) 262503.
\newblock

\bibitem{sen'kov10}
R.~Sen'kov, M.~Horoi, \prc\ 82 (2010) 024304.
\newblock

\bibitem{sen'kov11}
R.~Sen'kov, M.~Horoi, V.~Zelevinsky, \plettb\ 702 (2011) 413.
\newblock

\bibitem{sen'kov13}
R.~Sen'kov, M.~Horoi, V.~Zelevinsky, \cpc\ 184 (2013) 215.
\newblock

\bibitem{[Moe95]}
P.~M{\"o}ller, J.~R. Nix, W.~J. Myres, Swiatecki, \adndt\ 59 (1995) 185.

\bibitem{[Gor10]}
S.~Goriely, N.~Chamel, J.~M. Pearson, \prc\ 82 (2010) 035804.

\bibitem{[Ber05]}
G.~F. Bertsch, B.~Sabbey, M.~Uusn\"akki, \prc\ 71 (2005) 054311.

\bibitem{[Toi08]}
J.~Toivanen, J.~Dobaczewski, M.~Kortelainen, K.~Mizuyama, \prc\ 78 (2008)
  034306.

\bibitem{[Klu09]}
P.~Kl{\"u}pfel, P.-G. Reinhard, T.~J. B{\"u}rvenich, J.~A. Maruhn, \prc\ 79
  (2009) 034310.

\bibitem{Rei10}
P.-G. Reinhard, W.~Nazarewicz, \prc\ 81 (2010) 051303.
\newblock

\bibitem{more2010ecn}
J.~J. Mor{\'e}, S.~M. Wild, \sisc\ 33 (2011) 1292.
\newblock

\bibitem{more2011edn}
J.~J. Mor\'e, S.~M. Wild, \toms\ 38 (2012) 19:1.
\newblock

\bibitem{incite-home}
{U.S. Department of Energy}, {DOE Leadership Computing}, see
  \url{www.doeleadershipcomputing.org}.

\bibitem{NUCLEI}
{NUCLEI: Nuclear} computational low-energy initiative, see
  \url{http://computingnuclei.org}.

\end{thebibliography}
